\documentclass[journal, onecolumn, draftclsnofoot]{IEEEtran}

\usepackage{xr}
\usepackage{amsthm}
\usepackage[mathscr]{eucal}
\usepackage{amsfonts}
\usepackage[cmex10]{amsmath}
\usepackage{amssymb}
\usepackage{graphicx}
\usepackage{caption}
\usepackage{subcaption}
\usepackage{pstricks}
\usepackage{hyperref}
\usepackage{booktabs}
\usepackage{colortbl}
\usepackage{comment}
\usepackage{tabularx}

\usepackage{xcolor}

\usepackage[authoryear,round]{natbib}

\newcolumntype{N}{>{\centering\arraybackslash}m{.85in}}

\newrgbcolor{myblue}{0 0 1}
\newrgbcolor{myred}{0.466667 0 0}
\newrgbcolor{mygreen}{0 0.466667 0}
\newrgbcolor{mylightred}{1 0 0}
\newrgbcolor{mylightblue}{0 0.8 1}

\newcommand{\ttlred}[1]{\textcolor{mylightred}{#1}}

\usepackage{tabularx,ragged2e,booktabs,caption}
\newcolumntype{C}[1]{>{\Centering}m{#1}}

\newcommand\solidrule[1][0.5cm]{\rule[0.5ex]{#1}{.8pt}}

\usepackage{setspace}
\newcommand{\mb}[1]{\mathbb{#1}}

\newcommand{\bs}[1]{\boldsymbol{#1}}

\newcommand{\msf}[1]{\mathsf{#1}}

\newcommand{\tn}[1]{\textnormal{#1}}

\newcommand{\ind}{{1\hspace{-2.5pt}\tn{l}}}

\DeclareMathOperator*{\argmin}{argmin}

\newcommand{\R}{\mb{R}}

\newcommand{\N}{\mb{N}}
\newcommand{\E}{\mb{E}}

\newcommand{\SD}{\msf{SD}}
\newcommand{\MSE}{\msf{MSE}}
\newcommand{\COV}{\msf{Cov}}
\newcommand{\BIAS}{\msf{BIAS}}

\newcommand{\kernel}{F}
\newcommand{\sr}{f_{\bs{s}}}

\newcommand{\JULES}{\operatorname{JULES}}
\newcommand{\HSMUCE}{\operatorname{HSMUCE}}
\newcommand{\HMM}{\operatorname{HMM}}
\newcommand{\JSMURF}{\operatorname{JSMURF}}
\newcommand{\TRANSIT}{\operatorname{TRANSIT}}

\newcommand{\HILDE}{\operatorname{HILDE}}
\newcommand{\valmu}{c}

\usepackage{siunitx}
\DeclareSIUnit{\Molar}{\textsc{m}}
\usepackage{algorithm2e}
\usepackage{tikz}
\usetikzlibrary{shapes,arrows,backgrounds,fit,calc,positioning}

\numberwithin{equation}{section}
\newcounter{satze}
\numberwithin{satze}{section}
\theoremstyle{plain}

\newtheorem{Theorem}[satze]{Theorem}

\begin{document}

\title{Heterogeneous Idealization of Ion Channel Recordings - Open Channel Noise}

\author{Florian~Pein, Annika Bartsch, Claudia~Steinem and Axel~Munk%
\thanks{Florian Pein is with the Statistical Laboratory of the Department of Pure Mathematics and Mathematical Statistics (DPMMS) at the University of Cambridge, Wilberforce Road, Cambridge, CB3 0WB, United Kingdom.}%
\thanks{Annika Bartsch and Claudia Steinem  are with the Institute of Organic and Biomolecular Chemistry, Georg-August University of Goettingen, Tammannstr.~2, 37077 G\"ottingen, Germany.}%
\thanks{Axel Munk is with the Institute for Mathematical Stochastics, Georg-August-University of Goettingen, Goldschmidtstr. 7, 37077 G\"ottingen, Germany. Axel Munk is also with the Max Planck Institute for Biophysical Chemistry, Am Fassberg 11, 37077 G\"ottingen, Germany, and with the Felix Bernstein Institute for Mathematical Statistics in the Biosciences, Goldschmidtstr. 7, 37077 G\"ottingen, Germany.}%
\thanks{Support of DFG (CRC803, projects Z02 and A01, Cluster of excellence 2067 MBExC Multiscale Bioimaging: From Molecular Machines to Networks of Excitable Cells), the Volkswagen Foundation (FBMS) and of EPSRC (EP/N031938/1 - Statscale programme) is gratefully acknowledged. We thank Timo~Aspelmeier, Manuel~Diehn, Benjamin~Eltzner, Ingo~P.~Mey, Ole~M.~Sch\"utte, Ivo~Siekmann, Inder~Tecuapetla-G\'omez and Frank~Werner for fruitful discussions.
}}%

\maketitle

\begin{abstract}
We propose a new model-free segmentation method for idealizing ion channel recordings. This method is designed to deal with heterogeneity of measurement errors. This in particular applies to open channel noise which, in general, is particularly difficult to cope with for model-free approaches. Our methodology is able to deal with lowpass filtered data which provides a further computational challenge. To this end we propose a multiresolution testing approach, combined with local deconvolution to resolve the lowpass filter. Simulations and statistical theory confirm that the proposed idealization recovers the underlying signal very accurately at presence of heterogeneous noise, even when events are shorter than the filter length. The method is compared to existing approaches in computer experiments and on real data. We find that it is the only one which allows to identify openings of the PorB porine at two different temporal scales. An implementation is available as an \texttt{R} package.
\end{abstract}


\IEEEpeerreviewmaketitle

\section{Introduction}\label{sec:introduction}
The voltage patch clamp technique is a major tool to quantify the electrophysiological dynamics of ion channels in the cell membrane \citep{Neher.Sakmann.76, Sakmann.Neher.95}. It allows to record the conductance trace (i.e., the recorded current trace divided by the applied voltage) of a \textit{single} ion channel in time, which is for instance important in medical research for the development of new drugs \citep{kass2005channelopathies, overington2006many}. Important channel characteristics such as amplitudes and dwell times can be obtained provided the conductance changes of the traces are \textit{idealized} (underlying signal is reconstructed) from these recordings (data points) \citep{Colquhoun.87, Sakmann.Neher.95, Hotz.etal.13, Peinetal18}. To obtain such an idealization an extensive amount of methodology is available nowadays, a selective review is given below.

\paragraph*{Open channel noise}
In this paper, we focus on recordings that are affected by \textit{open channel noise}, i.e., have larger noise on segments with a larger conductance. The name open channel noise refers to the fact that a larger conductance results from an open pore. The additional noise when the channel is open can for instance be explained by current interruptions lasting approximately 1 microsecond \citep{sigworth1985open, sigworth1986open, sigworth1987open, heinemann1988open, heinemann1990open, heinemann1991open}. We analyze these recordings in a 'model-free' manner, i.e., without assuming a hidden Markov or related model, as a time series which is obtained by equidistant sampling from the convolution of a piecewise constant signal contaminated by white noise with the kernel of a lowpass filter. The white noise is scaled by an unknown piecewise constant standard deviation function to allow variance heterogeneity caused by open channel noise. We stress, that this modeling is rather general and also allows to deal with heterogeneity of measurement errors, not necessarily due to open channel noise. It will be explained in full detail in Section \ref{sec:model}.

\begin{figure}[!htb]
\centering
\includegraphics[width = 0.99\textwidth]{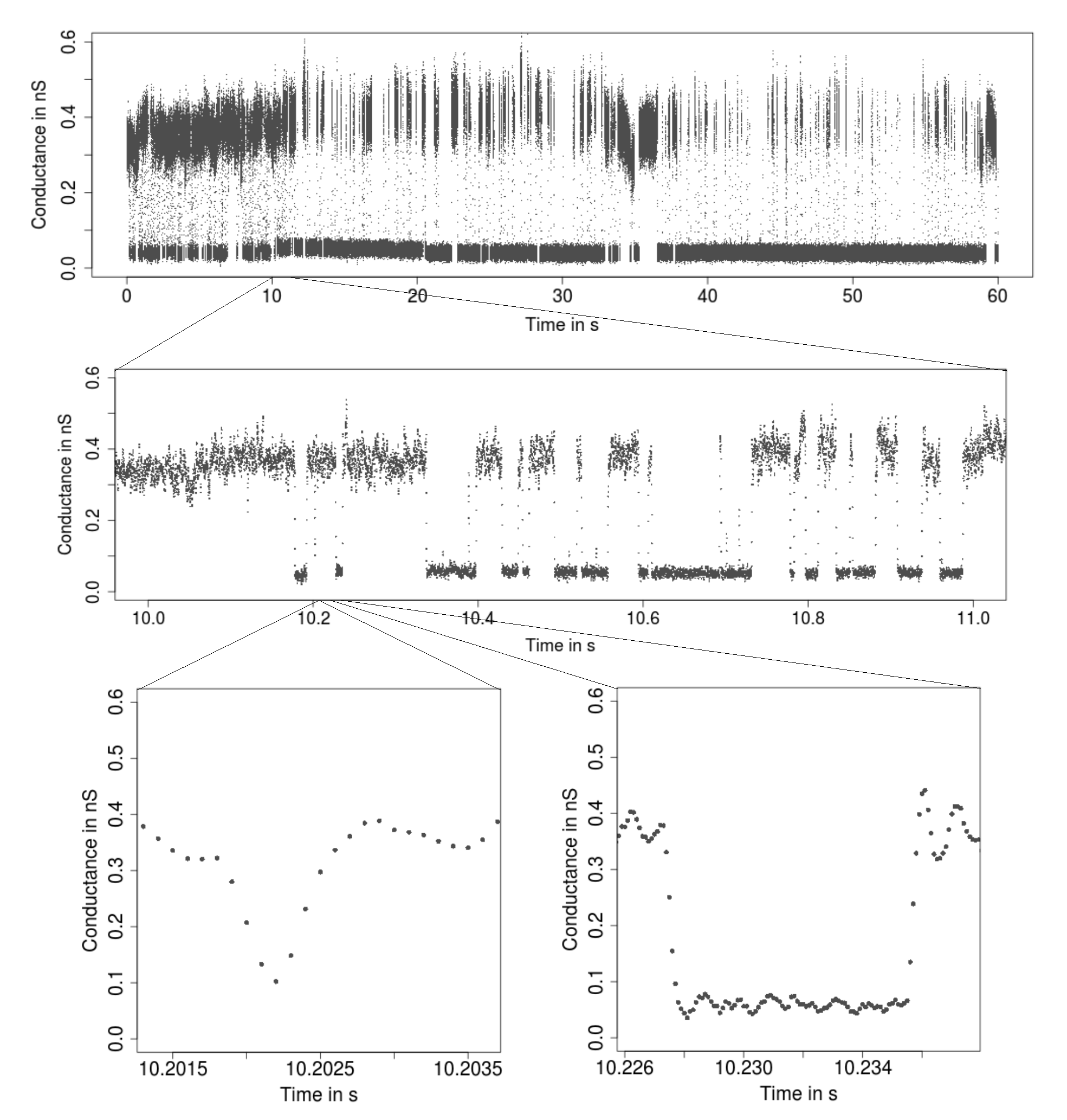}
\caption{From seconds to microseconds: Ion channel recordings (grey points) displayed at a level of seconds (top panel), of milliseconds (middle panel) and of microseconds (bottom panels). Data points result from a representative conductance recording of PorB by the patch clamp technique using solvent-free bilayers at $\SI{20}{\milli\volt}$.}
\label{fig:PorBHeteroData}
\end{figure}

\paragraph*{Data: the outer membrane porin PorB}
Figure \ref{fig:PorBHeteroData} shows exemplarily a conductance trace of the  outer membrane porin PorB from \textit{Neisseria meningitidis}, a pathogenic bacterium in the human nose and throat region \citep{virji2009pathogenic}. PorB is a trimeric porin and the second most abundant protein in the outer membrane of \textit{Neisseria meningitidis}. It is for instance relevant for the transport of antibiotics into the cell and hence of current interest to understand antibiotic resistance better. Recordings are obtained by the patch clamp technique using solvent-free bilayers. In Figure \ref{fig:PorBHeteroData} it is clearly visible that the two conductance levels around \SI{0.04}{\nano\siemens} and \SI{0.36}{\nano\siemens} are affected from open channel noise as the variance of observations around \SI{0.36}{\nano\siemens} is much larger than of the ones around \SI{0.04}{\nano\siemens}. Such recordings were manually analyzed in \citep{Bartschetal18} (Figure 1 and its explanation). They have been a major motivation for our work as they show distinct heterogeneous noise but also short event times, which we could not tackle satisfactorily by existing idealization methods, but also not by a manual analysis. In fact, in \citep{Bartschetal18} only the conductance levels were investigated but not the full gating dynamics, since events on short time scales could not be idealized.

\begin{figure}[!htb]
\centering
\includegraphics[width = 0.99\textwidth]{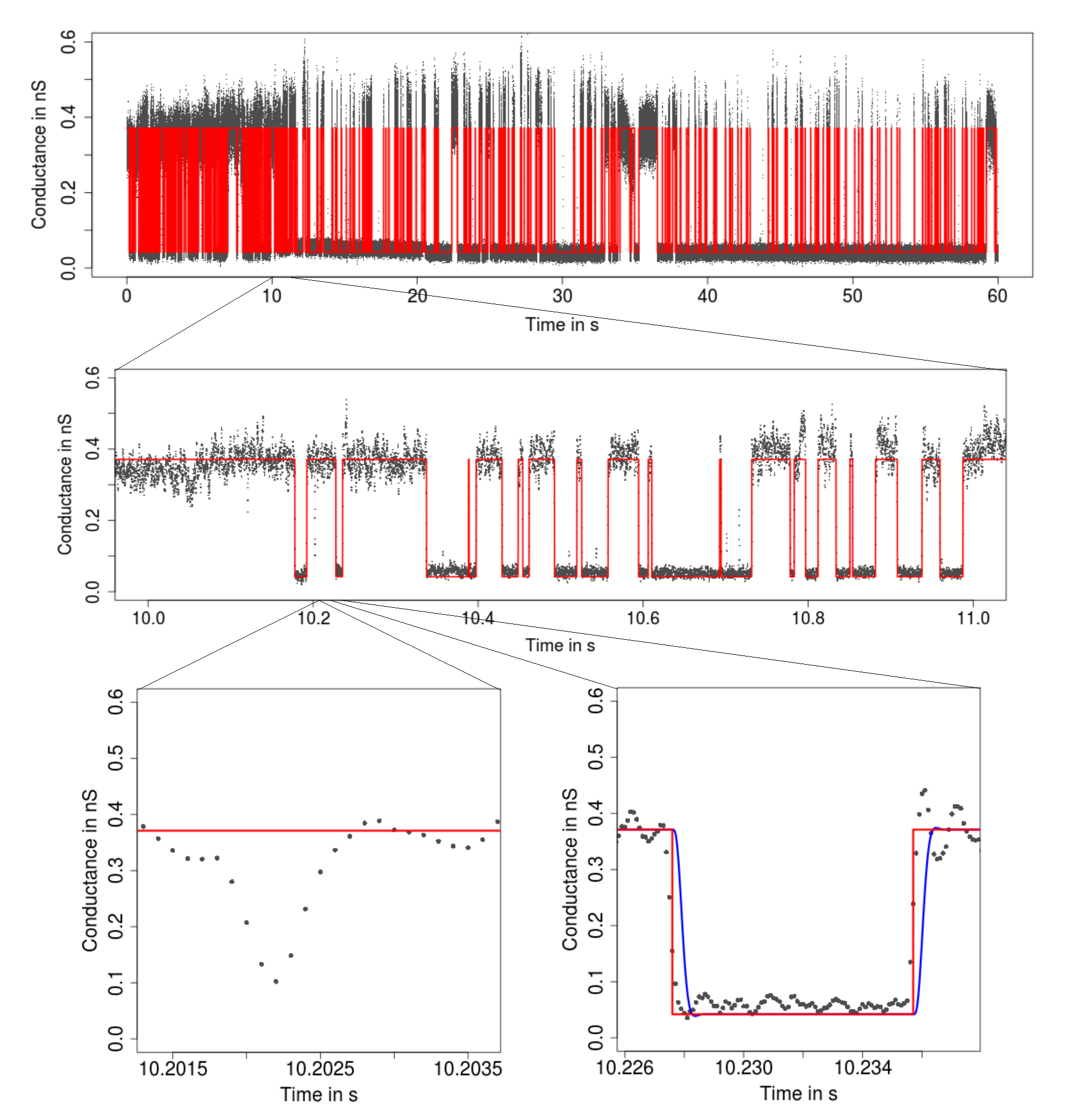}
\caption{Idealization (red) of the data in Figure \ref{fig:PorBHeteroData} assuming a HMM displayed on three different temporal scales. The model consists of three states, whereby two states are assumed to have the same expectation and variance. In the lower panels we also show the convolution of the idealization with the lowpass filter (blue). It fits most part of the data well, but misses short events, for instance the event displayed in lower left panel.}
\label{fig:PorBHeteroCoupled}
\end{figure}

\paragraph*{Methods for open channel noise}
Idealization methodology can be divided into so called model-free methods \citep{Colquhoun.87, VanDongen.96, Hotz.etal.13, gnanasambandam2017unsupervised, Peinetal18} which do not rely on a specific model for the gating dynamics, to methodology based on hidden Markov models (HMM) \citep{Ball.Rice.92, Venkataramanan.etal.00, Qin.etal.00, deGunst.etal.01, Siekmann.etal.11, diehn2019maximum} and to current distribution fitting \citep{yellen1984ionic, heinemann1991open, schroeder2015resolve, hartel2019high}. The latter often assume a hidden Markov model as well but focus on parameter estimation directly. An idealization can be obtained by the Viterbi algorithmus \citep{viterbi1967error} as soon as the parameters are determined.\\
Most HMM methods can deal with heterogeneous noise. Moreover, they allow to extrapolate information from larger (observable) to smaller (not observable) time scales and hence can provide a good idealization on small temporal scales. However, they rely heavily on the correctness of the assumed model assumptions. Up to few exceptions, see \citep{Fulinski.98, Goychuk.etal.05, Mercik.Weron.01, Shelley.etal.10}, a Markov model is a reasonable assumption for the underlying ion channel dynamics. However, artifacts in the data observed, for instance base line fluctuations, occur frequently in ion channel recordings and require elaborate data cleaning before a HMM can be fitted.  Base line fluctuations are for instance caused by small defects in the membrane, which is unavoidable in the recordings. There might be also periodic oscillations, resulting from the electronics or from building vibrations (although damped). The PorB measurements display in Figure \ref{fig:PorBHeteroData} show several artifacts of this type (see for instance the waviness of the observations or the conductance increase around \SI{10.1}{\second}, which severely hinders straightforward fitting by a HMM: We tried to fit this data set with in total four different hidden Markov model approaches. We achieved the best results when we assumed three states, but with the assumption that two states (with small conductivity) share the same expectation and variance. More details, also on parameter choices, are given in Section \ref{sec:appendixexamples} in the supplement. The obtained idealization is shown in Figure \ref{fig:PorBHeteroCoupled}. It fits long events well, but misses very short events, see for instance the lower left panel. Fitting such events well requires to take into account the filtering which is computationally very demanding for HMMs. We will discuss such an approach in Section \ref{sec:appendixexamples} in the supplement as well. In summary, in addition, to the low robustness against artifacts, the choice of a specific Markov model, especially the determination of the number of states, can be a demanding task and often involves subjective choices by the analyst.

Contrary, model-free approaches can deal way more flexible with artifacts as they act rather locally on the time series without the assumption of a global model. Hence they are more robust than HMMs to model violations. Therefore, they complement HMMs well, e.g., as a preprocessing step. For example, model-free methods can be used to select or verify a specific Markov model, in particular to determine the number of states and possible transitions, as they explore and potentially remove artifacts in a model-free manner. See also \citep{Peinetal18} for a more extensive discussion of further aspects of the different approaches.\\
To the best of our knowledge, all existing model-free approaches assume (implicitly or explicitly) homogeneous noise and hence produce unreliable results when open channel noise is present. Among the first methods which fall into this category is $\TRANSIT$ \citep{VanDongen.96}. An idealization by this approach, details of its limitations in our setting and further discussions can be found in Section \ref{sec:appendixexamples} in the supplement. In Figure \ref{fig:PorBHeteroJULES} we display $\JULES$ \citep{Peinetal18}, a novel multiscale approach that also falls into this type of methods. It detects many small events on segments with a larger conductance and variance, but none on ones with a smaller conductance and variance. These additional events are most likely artifacts caused by open channel noise. Indeed, in Section \ref{sec:hmm} we found that the rates of a simulated hidden Markov model with parameters similar to them underlying the observations in Figure \ref{fig:PorBHeteroData} could not be recovered when we used $\JULES$ to idealize the underlying signal. This effect is even more severe when the variance heterogeneity is larger. 

\begin{figure}[!htb]
\centering
\includegraphics[width = 0.99\textwidth]{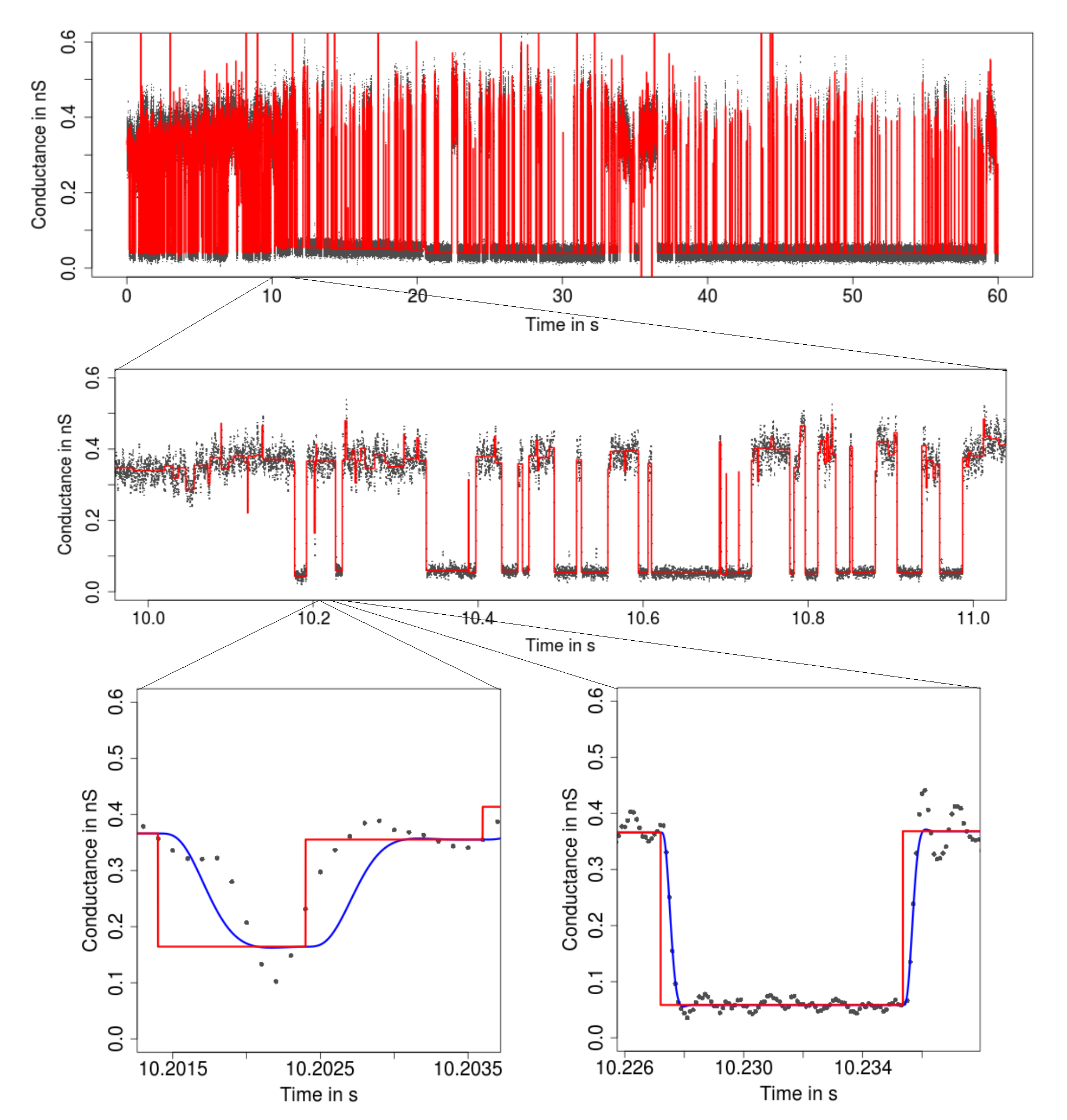}
\caption{Idealization (red) of the data in Figure \ref{fig:PorBHeteroData} by $\JULES$ displayed on three different temporal scales. In the lower panels we also show the convolution of the idealization with the lowpass filter (blue). It detects short events, but finds many small events (which are most likely false positives) at parts of high conductance and high variance (see for instance the idealization of the observations around \SI{0.36}{\nano\siemens} in the middle panel). These detections hinder the decovolution (see for instance the lower left panel) and make the idealization unreliable.}
\label{fig:PorBHeteroJULES}
\end{figure}

\begin{figure}[!htb]
\centering
\includegraphics[width = 0.99\textwidth]{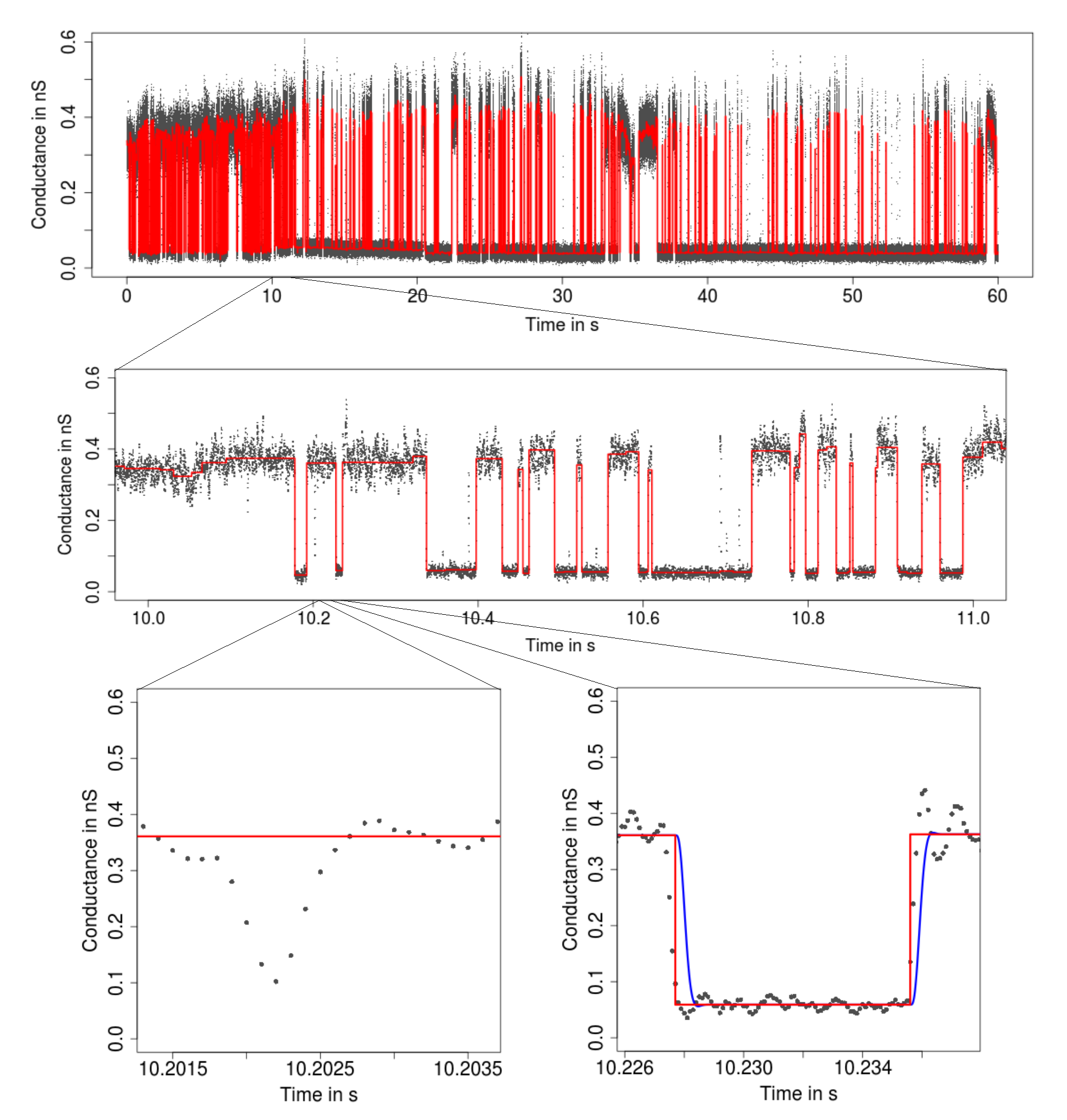}
\caption{Idealization (red) of the data in Figure \ref{fig:PorBHeteroData} by $\operatorname{HSMUCE}$ displayed on three different temporal scales. In the lower panels we also show the convolution of the idealization with the lowpass filter (blue). It detects events on larger temporal scales well (middle panel, lower right panel), but misses short events, see for instance at around \SI{10.2}{\second} (lower left panel), \SI{10.4}{\second} or \SI{10.7}{\second}.}
\label{fig:PorBHeteroHSMUCE}
\end{figure}

Recently, there has been made some progress to adjust for heterogeneous noise in the context of model-free methods. However, they are not dedicated to idealize ion channel recordings, which means in particular that they do not incorporate lowpass filtering. Obviously, ignoring the filtering will deteriorate results. For illustration purposes we display the heterogeneous multiscale approach $\operatorname{HSMUCE}$ \citep{Pein.Sieling.Munk.16} in Figure \ref{fig:PorBHeteroHSMUCE}. We found that it provides reasonable results on larger temporal scales. However, due to filtering $\operatorname{HSMUCE}$ misses shorter events, see for instance the missed peaks around \SI{10.2}{\second} (lower left panel), \SI{10.4}{\second} or \SI{10.7}{\second}. Also for this type of methods we provide further examples and discussions in Section \ref{sec:appendixexamples} in the supplement.

The occurrence of short events is often called \textit{flickering}. Missing them does not only potentially disturb the analysis of the general channel behavior, the analysis and hence the idealization of flickering events is also of its own interest in many applications, since flickering has often its own dynamics and can result from different molecular processes. Typical examples are conformational changes of the ion channel \citep{Grosse.etal.14} or the passage of larger molecules blocking the ions pathway through the channel \citep{Singh.etal.12, Bartschetal18}. Hence, one main goal of this paper will be to idealize and detect such events as well.\\
To this end, we introduce in Section \ref{sec:model} a statistical model which resembles all features (open channel noise, events on a large range of scales, filtering) of such complex data as in the previous example. In summary, we then ask for a model-free idealization method that adapts automatically to heterogeneous noise, hence is able to detect and idealize events on a large range of relevant scales accurately, but in particular also events shorter than the filter length. Furthermore, we aim to provide theoretical justification for the detected events (controlling false positives) and for a computationally efficient method to deal with large data sets.

\paragraph*{HILDE}
To address these tasks, we propose in this paper a new method called \textbf{H}eterogeneous \textbf{I}dealization by \textbf{L}ocal testing and \textbf{DE}convolution, $\HILDE$. This method combines multiscale regression and deconvolution as it takes into account the convolution of the signal with the lowpass filter explicitly for detecting events that are short in time. Before we will explain our (quite involved) methodology further, we discuss firstly the general challenges: A major difficulty for any such method due to the presence of heterogeneous noise is to distinguish between small jumps in the signal and random fluctuations caused by the noise of unknown level. However, simultaneously estimating the signal and the noise level locally is notoriously difficult in general \citep{Pein.Sieling.Munk.16} and further hampered in our situation since the unknown signal and noise are both smoothed by the filter and hence deconvolution is required when shorter temporal scales are considered. We solve this by means of a multiresolution approach in combination with a local deconvolution to idealize events on all relevant temporal scales accurately. Whereas statistical multiresolution idealization that ignores the deconvolution can be computed efficiently by dynamic programming, see for instance \citep{Hotz.etal.13, Frick.Munk.Sieling.14, Pein.Sieling.Munk.16}, combining multiresolution procedures with deconvolution is algorithmically difficult, since due to the coupling of all observations in the idealization, dynamic programming is not applicable without further ado. We will overcome this burden by focusing firstly on larger temporal scales and then improving the idealization on smaller temporal scales. More precisely, $\HILDE$ consists of the following three steps: a) detection of long events, b) detection of short events and c) parameter estimation by deconvolution. A summary about all three steps is given in Algorithm \ref{alg:hilde} (see Section \ref{sec:methodology}).

\paragraph{Detection of long events} We will obtain an idealization by multiresolution regression that covers all important features on larger temporal scales (for the data set analyzed here large means events of length \SI{6.5}{\milli\second} at least, i.e., $\geq 65$ sampling points). This step is discussed in Section \ref{sec:large scales} and technical details are given in Section \ref{sec:appendixlongsegments} in the supplement.

\begin{figure}[!htb]
\centering
\includegraphics[width = 0.99\textwidth]{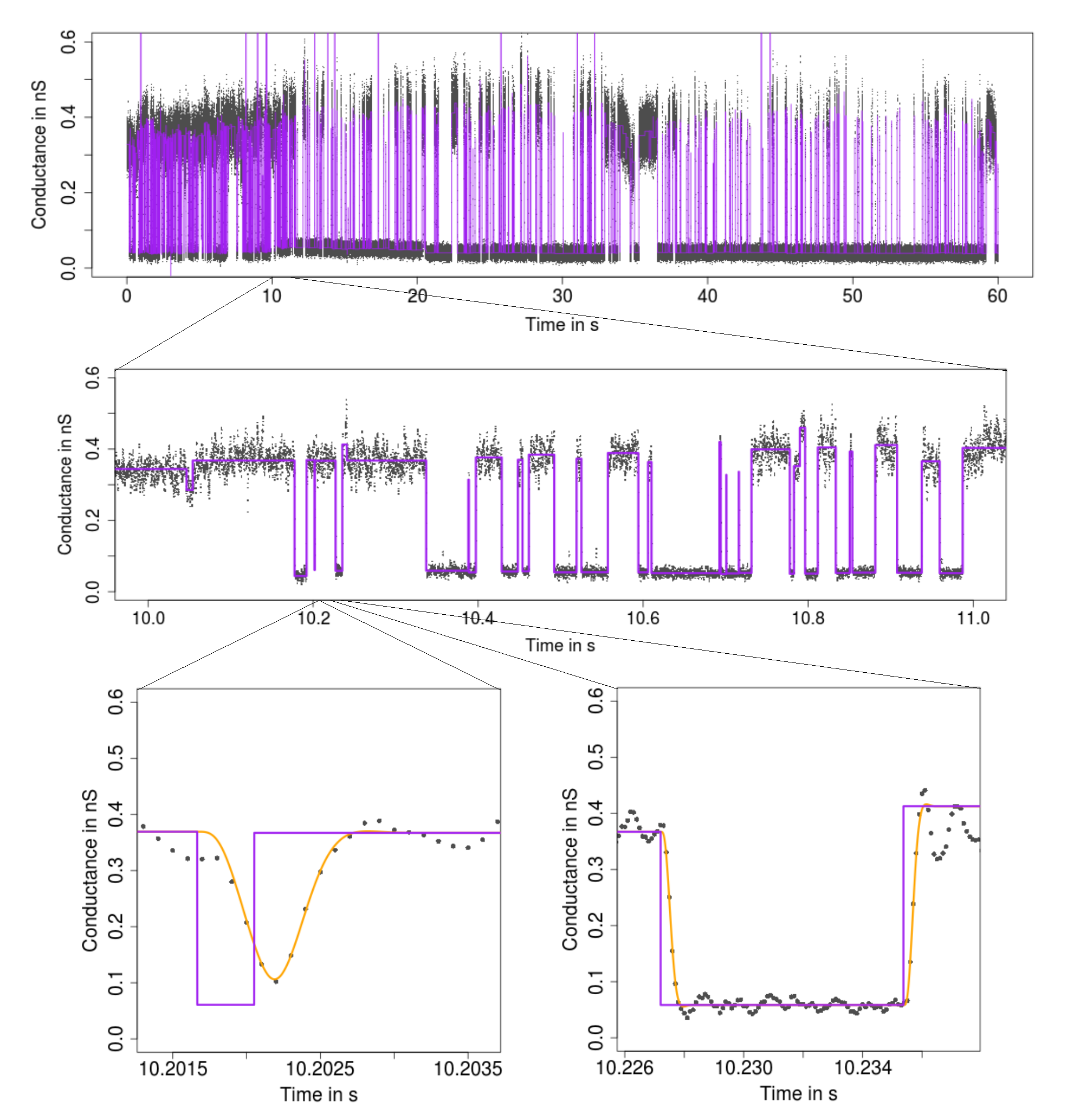}
\caption{Idealization (purple) of the data in Figure \ref{fig:PorBHeteroData} by $\HILDE$ displayed on three different temporal scales. In lower panels we also show the convolution of the idealization with the lowpass filter (orange). $\HILDE$ idealizes events on all relevant temporal scales well.}
\label{fig:PorBHeteroJILTAD}
\end{figure}

\paragraph{Detection of short events}
Our data set contains several short events that will be missed by the previous step, see for instance in Figure \ref{fig:PorBHeteroData} at around \SI{10.2}{\second} (lower left panel), \SI{10.5}{\second}, \SI{10.7}{\second} and \SI{10.8}{\second}. To detect such events, we test locally whether additional events on smaller temporal scales have to be incorporated. This is impaired by the lowpass filter and the resulting convolution has to be taken into account explicitly. To this end, we assume that signal and noise left and right of the interval on which we test are given by the idealization from the previous multiresolution step. These tests are detailed in Section \ref{sec:smallscales}, while technical details are postponed to Section \ref{sec:appendixshortsegments} in the supplement.\\
Steps a) and b) determine the number of events and their rough locations. The final idealization in Figure \ref{fig:PorBHeteroJILTAD} (see e.g. the lower left panel) confirms that step b) is indeed able to detect short events (up to \SI{0.2}{\milli\second}, corresponding to only two subsequent observations).
 
\paragraph{Parameter estimation by deconvolution} 
Finally, the precise locations of the events and the conductance levels have to be obtained. This will done in an additional deconvolution step, as the recordings are filtered. To this end, we use the local deconvolution approach from \citet{Peinetal18} with minor modifications. This step is discussed in Section \ref{sec:deconvolution} and technical details are explained in Section \ref{sec:appendixdeconvolution} in the supplement.\\
Figure \ref{fig:PorBHeteroJILTAD} shows the final idealization by $\HILDE$ of the observations in Figure \ref{fig:PorBHeteroData}. Despite distinct heterogeneous noise, the idealization covers all main features on all relevant scales, in particular also short events, while at the same time it does not include systematically additional artificial changes. The zooms into single peaks (lower panels) show that $\HILDE$ fits the observations well down to a scale of microseconds, which is also a confirmation of our approach, including the modeling. We stress that $\HILDE$ is not only robust against heterogeneous noise but has typically also a larger detection power for event detection than $\JULES$ (even when the noise is homogeneous), since it takes into account the convolution explicitly for detection. This is discussed in more detail in Section \ref{sec:discussionhomogeneous}, where we also outline a version of $\HILDE$ that assumes homogeneous noise to improve detection power even further if the homogeneous noise assumption is justified.

While the first and the third step are mostly useful modifications of existing methodologies, we want to stress that this is not true for the second step. To the best of our knowledge, no other model-free ion channel idealization method is able to take the convolution explicitly into account when detecting events. As discussed before, this is however indispensable to detect short events when filtering and heterogeneous noise are present.

\paragraph*{Implementation and run time}
Each step of $\HILDE$ can be computed separately. Hence, $\HILDE$ can be applied and modified in modular fashion. This allows for instance to skip the second step if a data set contains only longer events, hence saving computation time. Another usage might be to modify the local tests in the second step, for instance to increase the detection power in a data set with small conductance changes but large difference in the noise levels, without modifying the first or third step. Such modifications are discussed in Section \ref{sec:discussionalternatives}.\\
The first multiresolution regression step can be computed by a pruned dynamic program. The computation of the local tests in the second step is straightforward and the deconvolution in the third step can be computed by an iterative grid search. These steps are detailed in Section \ref{sec:computation} and summarized in Algorithm \ref{alg:hilde}. An implementation is available by the function \textit{hilde} in the \texttt{R} package \textit{clampSeg} accompanying this paper. The package is available on request and has been submitted parallel to CRAN \citep{clampSeg}.\\
The worst case computational complexity is quadratic in the number of observations, but in most ion channel recordings conductance changes occur frequently which reduces the complexity to linear in the number of observations. For instance the $600\,000$ observations in Figure \ref{fig:PorBHeteroData} can be idealized in a few minutes on a standard laptop. A detailed discussion of the computational complexity is given in Section \ref{sec:computation}.

\paragraph*{Simulations}
In Section \ref{sec:simulations} we investigate the performance of $\HILDE$ in Monte-Carlo simulations which resemble the characteristics of the data in the application in Section \ref{sec:analysis}. Based on this we confirm that $\HILDE$ works very well for data sets like the one shown in Figure \ref{fig:PorBHeteroData}. In more detail, it can detect events which last \SI{0.2}{\milli\second}, corresponding to only two subsequent observations and being less than one fifth of the filter length long, with probability almost one. Furthermore, all parameters (conductance levels and the locations of the changes) are estimated very accurately, see Section \ref{sec:singlePeak} for more details. Moreover, two subsequent events can be separated reliably as soon as the distance between them is larger than five times the filter length, see Section \ref{sec:consecutivePeaks}. In Section \ref{sec:hmm} we simulate data from a hidden Markov model. Our method is not assuming a HMM, but such a model is still illustrative to simulate as it is a standard assumption for the analysis of ion channel recordings. We will also see in Section \ref{sec:analysisGating} that a Markov model is reasonable for the PorB recordings. We find in Section \ref{sec:hmm} that $\HILDE$ recovers all parameters with high precision. Those parameters are chosen similar to those which we have estimated in Section \ref{sec:analysisGating}. Finally, we investigate robustness issues against $f^2$- and $1/f$-noise in Section \ref{sec:robustness}. We omit most of the time a systematic comparison with other approaches, since, as discussed in the introduction before, to the best of our knowledge all existing approaches assume a more restrictive model which hinders a fair comparison. However, we include $\JULES$, $\HSMUCE$ and a $\HMM$ based approach in the simulations in Section \ref{sec:hmm} to illustrate the shortcomings (and benefits) of these approaches further.

\paragraph*{Application to PorB recordings}
Our analysis of single channel recordings of PorB in Section \ref{sec:analysis} confirms all major results from \citep{Bartschetal18} about this data set. Moreover, a novel finding of our analysis is that the dwell times do not fit a single exponential distribution, but suggests that two different regimes for the dwell times are underlying: very short openings of \SI{2.31}{\milli\second} estimated average duration and longer openings of \SI{51.62}{\milli\second} estimated average duration.  To best of our knowledge, fast and slow gating at the same time was not observed for PorB before, but for another porine OmpG \citep{Grosse.etal.14}. We stress that all results obtained by $\HILDE$ could be confirmed by at least one other approach. However, none of the other methods were able to reproduce all results obtained by $\HILDE$.\\
In summary, in this work we proposed with $\HILDE$ the first fully automatic model-free method for the analysis of ion channel recordings affected from open channel noise, i.e., to the best of our knowledge no other existing methodology is able to estimate a piecewise constant function in a model-free manner when filtering and heterogeneous noise are present at the same time. Simulations confirm that $\HILDE$ deals efficiently with heterogeneous noise and filtered data at the same time and idealizes events on various time scales efficiently. More precisely, to obtain a good idealization events have to be only at least two subsequent observations long but separated from each other by at least five times the filter length (at signal and noise ratio and filtering as in the analyzed data). This allowed us to obtain novel findings for the PorB channel, e.g., that it can have shorter and longer opening processes at the same time. 

\section{Modeling}\label{sec:model}
We assume that the recordings result from equidistant sampling from the convolution of an unknown piecewise constant signal corrupted by Gaussian white noise with the (known) kernel of a lowpass filter. We stress however that our methodology can be extended to an unknown filter by using the methodology of \citep{Munk.Tecuapetla.15}. To incorporate heterogeneity, the white noise is scaled by an unknown piecewise constant function to allow a larger variance on segments on which the conductance is larger. We only allow potential variance changes when the conductance changes, since variance changes also depend on gating events of the channel. More precisely, we model the conductivity and the standard deviation by piecewise constant signals $f$ and $\sigma$,
\begin{equation}\label{eq:signal}
    f(t) = \sum_{j=0}^{K}\,\valmu_j\,\ind_{[\tau_j, \tau_{j+1})}(t) \text{ and } \sigma(t) = \sum_{k=0}^{K}\,s_k\,\ind_{[\tau_k, \tau_{k+1})}(t),
\end{equation}
where $t$ denotes physical time. The (unknown) conductance levels are denoted as $\valmu_0, \ldots, \valmu_K$, the (unknown) standard deviations as $s_0,\ldots,s_K>0$, the (unknown) number of changes as $K$ and the (unknown) locations as $-\infty =: \tau_0 < \tau_1 < \cdots < \tau_K < \tau_{K+1}:=\tau_{\operatorname{end}}$. The indicator function $\ind_A(t)$ is one if $t\in A$ and zero otherwise. The signals are extended to $\tau_0=-\infty$ to define the convolution correctly but we will see at the end of this section that only a very short time period before recordings started, i.e., before $t=0$, will be relevant. We assume $\valmu_k\neq \valmu_{k+1}$ to define the number of changes unambiguously, i.e., to obtain an \textit{identifiable} model. But we allow $s_k=s_{k+1}$, i.e., the standard deviation does not have to change between different events and in particular homogeneous noise is still part of the model ($\sigma \equiv s_0$). We stress that the class of signals in \eqref{eq:signal} is very flexible as potentially any arbitrary number of changes at arbitrary conductance levels and arbitrary standard deviations can be imposed, see Figure \ref{fig:PorBHeteroJILTAD} for an example.\\
We assume further that the recorded data points $Y_1,\ldots,Y_n$ (the measured conductivity at time points $t_i=i/\sr,\ i = 1,\ldots,n$, equidistantly sampled at rate $\sr$) result from convolving the signal $f$ perturbed by Gaussian white noise $\eta$ scaled by the standard deviation function $\sigma$ with an analogue lowpass filter, with (truncated) kernel $\kernel_m$, and digitization at sampling rate $\sr = n / \tau_{\operatorname{end}}$, i.e.,
\begin{equation}\label{eq:model}
Y_i = \big(\kernel_m \ast (f + \sigma \eta)\big)\left(i/\sr\right) = (\kernel_m \ast f)(i/\sr) + \epsilon_i,\quad i=1,\ldots,n,
\end{equation}
with $\ast$ the convolution operator. Here, $n$ denotes the total number of data points (typically several hundred thousands up to few millions). Like in \citep{Hotz.etal.13, Peinetal18} we truncate (and rescale) the kernel of the lowpass filter and the covariance function at $m/ \sr$ to simplify our model. This is implemented in the \texttt{R} function \textit{lowpassFilter} \citep{clampSeg}. 
As a working rule, we choose $m$ such that the autocorrelation function of the untruncated analogue lowpass filter is below $10^{-3}$ afterwards. For the later analyzed PorB traces, which are filtered by a 4-pole lowpass Bessel filter with $\SI{1}{\kilo\hertz}$ cut-off frequency and sampled at $\SI{10}{\kilo\hertz}$, this choice leads to $m = 11$. Hence, the resulting errors $\epsilon_1,\ldots,\epsilon_n$ are Gaussian and centered, $\E[\epsilon_i]=0$, and have covariance 
\begin{equation}\label{eq:cov}
\COV\big[Y_i,Y_{i+j}\big]=\left\{\begin{array}{cl}
\sum_{k=0}^{K}{s_k^2 \big[\mathcal{A}_m(i/\sr - \tau_{k}, j / \sr)-\mathcal{A}_m(i/\sr - \tau_{k+1}, j / \sr)\big]} & \text{for } \vert j \vert=0,\ldots,m,\\
0 & \text{for } \vert j\vert > m,
 \end{array}\right.
\end{equation}
with
\begin{equation}\label{eq:tacf}
\mathcal{A}_m(t,l):=\int_{0}^{t}{\kernel_m (s)\kernel_m(s + l)ds}.
\end{equation}
Note that we have in \eqref{eq:cov} an unknown non-stationary covariance structure. However, the covariance can be decomposed into a known stationary autocorrelation given by the lowpass filter and an unknown non-stationary variance, which is modeled by a piecewise constant function that shares its change-points with the mean function. An analytic expression of $\mathcal{A}_m(t,l)$ is implemented in the \textit{R} function \textit{lowpassFilter}. Hence, \eqref{eq:cov} can be computed exactly and efficiently.\\
The major aim will be now to idealize (reconstruct) the unknown signal $f$ taking into account the convolution, the heterogeneous noise given by \eqref{eq:cov} and the specific structure of $f$ in \eqref{eq:signal}. This will be done fully automatically and with statistically error control. By fully automatic we mean that no user action is required during the idealization process, only certain errors levels $\alpha=\alpha_1+\alpha_2$, the maximal scale on which local tests are performed $l_{\max}$ and two filter specific parameters have to be selected in advance, see Section \ref{sec:parameterchoices}.

\section{Methodology: HILDE}\label{sec:methodology}
In this section we detail the three steps of our \textbf{H}eterogeneous \textbf{I}dealization by \textbf{L}ocal testing and \textbf{DE}convolution ($\HILDE$) approach. A summary of these steps is given in the Meta-algorithm \ref{alg:hilde}.

\tikzstyle{line} = [draw, -latex']
\begin{algorithm}
\center
\scalebox{0.95}{
\begin{tikzpicture}[node distance = 1.5cm, auto] \centering
\node [draw,trapezium,trapezium left angle=70,trapezium right angle=-70,minimum width=12cm,fill=orange!20] (input) {\shortstack{Data $Y_1,\ldots,Y_n$, error levels $\alpha=\alpha_1+\alpha_2$, maximal scale $l_{\max}$,\\ regularization parameter $\gamma^2$, filter with kernel $\kernel_m$ truncated at $m / \sr$}};

\node[rectangle,fill=red!20, below of=input,minimum width=12cm] (reconstruction) {\shortstack{\textbf{Detection of long events} ($> l_{\max}$): \\ Multiresolution regression at error level $\alpha_1$}};

\node[rectangle,fill=red!20, below = 0.5cm of reconstruction,minimum width=12cm] (postfiltering) {\shortstack{\textbf{Detection of short events}: ($\leq l_{\max}$):\\ Local tests that take into account the convolution explixitly, at joint level $\alpha_2$}};

\node[rectangle,fill=red!20, below of=postfiltering,minimum width=12cm] (deconvolution) {\shortstack{\textbf{Parameter estimation}:\\
Local deconvolution, regularized by $\gamma^2$}};       

\node [draw,trapezium,trapezium left angle=70,trapezium right angle=-70,minimum width=12cm,fill=orange!20, below of=deconvolution] (output) {Idealization $\hat{f}$, i.e., all event times and conductance levels};

\path [line] (input) -- (reconstruction);
\path [line] (reconstruction) -- (postfiltering);
\path [line] (postfiltering) -- (deconvolution);
\path [line] (deconvolution) -- (output);

\end{tikzpicture}
}
\caption{Steps of $\HILDE$.}
\label{alg:hilde}
\end{algorithm}

\subsection{Detection of long events}\label{sec:large scales}
To detect events on larger temporal scales, we use a modification of the \textbf{H}eterogeneous \textbf{S}imulataneous \textbf{MU}ltiscale \textbf{C}hange-point \textbf{E}stimator, $\operatorname{HSMUCE}$ from \citep{Pein.Sieling.Munk.16}, which is a multiresolution procedure that is robust against heterogeneous noise. To avoid false positives due to the filter, we omit on each interval the first $m$ observations and do not test on very short intervals. Since we truncated the filter, the signal and the convolution of the signal with the lowpass filter differ only at the beginning of each segment. More precisely, if the signal is constant on an interval $[i/\sr, j/\sr]$ with conductance level $\valmu_{ij}$ and the first $m$ observations $Y_{i},\ldots,Y_{i+m-1}$ are ignored, all other observations $Y_{i+m},\ldots,Y_{j}$ have constant expectation equal to the conductance level $\valmu_{ij}$. Hence, we take into account only intervals longer than $m/\sr$ and ignore the first $m$ observations of each interval.\\
This leads to an estimator that detects change-points at presence of heterogeneous noise and filtering, i.e., when the heterogeneous ion channel model from Section \ref{sec:model} is assumed, while at the same time the probability to overestimate the number of events is controlled, i.e., a false positive is only added with probability at most equal to the tuning parameter $\alpha_1$, see Theorem \ref{theorem:overestimation} in Section \ref{sec:appendixlongsegments} in the supplement. A detailed definition of this estimator is given in Section \ref{sec:appendixlongsegments} in the supplement.
Note that it does not take into account the convolution explicitly but still has good detection properties if events are long enough, but almost no detection power on small scales. Simulations (not displayed) show that for our data set events with of length at least \SI{6.5}{\milli\second}, corresponding to $65$ sampling points, are detected reliably.\\
In the following two sections we will present a refinement of this idealization to detect and idealize events on smaller time scales, too, which proves to be relevant for our data example. Note that in this section and in the next section (a refinement will be provided in the local deconvolution step) we restrict all changes to the grid on which the observations are given, in other words, we assume that $\sr\tau_i$ are integers.

\subsection{Detection of short events}\label{sec:smallscales}
To detect short events, we test on all intervals containing $l=1,\ldots,l_{\max}$ (to be defined later) observations whether the previous idealization is the underlying signal or whether the inclusion of an additional event on the considered interval is significantly better. More precisely, let $[\tau_L,\tau_R]=[i/\sr,j/\sr]$ be the interval on which we test. And assume for the moment that $\tau$ is the only change in $[(i-m+1)/\sr, (j+m-1)/\sr]$ with conductance levels $\valmu_L$ before and $\valmu_R$ afterwards. Note that this also includes the scenario of no change in $[(i-m+1)/\sr, (j+m-1)/\sr]$ by setting $\valmu_L=\valmu_R$. Then, we decide whether an additional event on $[\tau_L,\tau_R]$ is required by testing the hypothesis 
\begin{equation}\label{eq:hypothesis}
f_0(t)=\begin{cases}
\valmu_L &\text{ if } t < \tau,\\
\valmu_R &\text{ if } t \geq \tau
\end{cases}
\end{equation}
against the alternative
\begin{equation}\label{eq:alternative}
f_1(\valmu)(t)=\begin{cases}
\valmu_L &\text{ if } t < \tau_L,\\
\valmu &\text{ if } \tau_L \leq t < \tau_R,\\
\valmu_R &\text{ if } t \geq \tau_R,
\end{cases}
\end{equation}
with $\valmu\in \R$ arbitrary. The same structure is assumed for standard deviation functions $\sigma_0$ and $\sigma_1$ with values $s_L, s$ and $s_R$. The precise hypotheses and alternatives, i.e., the values for $\tau, \valmu_L, \valmu_R, s_L$ and $s_R$, are determined by the previous idealization step. If more than one change is contained in $[(i-m+1)/\sr, (j+m-1)/\sr]$, no local test will be performed on this interval. The reasoning behind this and how to obtain $\tau, \valmu_L, \valmu_R, s_L$ and $s_R$ exactly are explained in the paragraph '\textit{Obtaining the hypotheses and alternatives}' in Section \ref{sec:appendixshortsegments} in the supplement. All tests are performed at simultaneous error level $\alpha_2>0$.\\
The form of these hypotheses allows us to construct a test statistic that takes into account the convolution explicitly. Moreover, information provided by potential variance changes can be used as well. We provide details of the corresponding test in the paragraph '\textit{Local testing}' in Section \ref{sec:appendixshortsegments} in the supplement. All choices there are motivated by a trade-off between a good detection power for events in the measurements in Section \ref{sec:analysis}, see Figure \ref{fig:PorBHeteroData}, and a reasonable computational complexity.\\
If a hypothesis is rejected, we replace the single change-point at $\tau$ by a short peak. Temporary locations will be placed at $\tau_L=i/\sr$ and $\tau_R=j/sr$, but exact locations and the conductance level $c$ will be obtained in the upcoming deconvolution step. However, note that usually one event in the data causes rejections of multiple tests. Therefore, we only consider the event with the largest test statistic among all rejections on intervals that intersect or adjoin each other. More details are provided in the paragraph '\textit{Multiple dependent rejections}' in Section \ref{sec:appendixshortsegments} in the supplement.

\subsection{Parameter estimation by local deconvolution}\label{sec:deconvolution}
The final idealization is obtained by local deconvolution as described in Section 3.2 of \citep{Peinetal18} with two adjustments, which will be discussed in Section \ref{sec:appendixdeconvolution} in the supplement. This means in particular that we still use the likelihood function of observations with homogeneous noise, although heterogeneous noise is assumed. Simulations show, see Section \ref{sec:simulations}, that this works reasonably well for the recordings we analyze in Section \ref{sec:analysis}. Alternatives for recordings with more pronounced noise heterogeneity are discussed in Section \ref{sec:discussionalternatives}.

\subsection{Computation and run time}\label{sec:computation}
The multiresolution regression step in Section \ref{sec:large scales} can be computed by a pruned dynamic program as described in Section A.1 in the supplement of \citep{Pein.Sieling.Munk.16}. For related ideas, see also \citep{killicketal12, Frick.Munk.Sieling.14, Lietal.16, maidstone2017optimal} and the references given there. The implementation of the local tests in Section \ref{sec:smallscales} is straightforward. The local deconvolution in Section \ref{sec:deconvolution} can be computed by an iterative grid search as described in Section 3.2 of \citep{Peinetal18}. An implementation of $\HILDE$ is available by the \texttt{R} function \textit{hilde} in the package \textit{clampSeg}. The package is available on request and has been submitted parallel to CRAN \citep{clampSeg}. All run time critical parts are written in C++ and are interfaced by the \texttt{R} code.\\
The worse case computation complexity of the dynamic program is quadratic in the number of observations $n$. However, in most ion channel recordings conductance changes occur frequently which reduces the complexity to be linear $\mathcal{O}(n)$, see Section A.3 in the supplement of \cite{Pein.Sieling.Munk.16}. The local tests in Section \ref{sec:smallscales} are of complexity $\mathcal{O}(l_{\max}^2 n)$, since for each of the $l_{\max}$ scales $1,\ldots,l_{\max}$ roughly $n$ tests have to be performed and the complexity to compute a single test is at most of order $\mathcal{O}(l_{\max})$. The computation time of the local deconvolution is dominated by the iterative grid search to deconvolve a single event. The deconvolution of a single event is constant in the number of observations, since the number of involved observations and the grid sizes do not increase. Moreover, the number of involved observations is small and the covariance matrix is a band matrix, with band size equal to $m$, which allows fast computation. Hence, the complexity of the deconvolution increases linearly in the number of events which increases for ion channel recordings typically linearly in the number of observations. In summary, for a typical channel trace the complexity to compute $\HILDE$ increases only linearly in the number of observations. This is confirmed by a run time of less than five minutes for idealizing the $600\,000$ observations in Figure \ref{fig:PorBHeteroData} on a Dell Latitude E6530 with Intel(R) Core(TM) i5-3340M CPU 2.70GHz processor. Similar run times are obtained for the traces generated in Section \ref{sec:hmm}. Thus, the theoretical considerations as well as the empirical run times confirm that $\HILDE$ can be computed efficiently, which is important since large data sets have to be analyzed.

\subsection{Parameter choices}\label{sec:parameterchoices}
$\HILDE$ can be tuned by the parameters $\alpha_1,\alpha_2, l_{\max}$ and $\gamma^2$, see Algorithm \ref{alg:hilde} and the referenced sections for a definition. The probability to overestimate the number of conductance changes is approximately controlled by the sum of the error levels $\alpha = \alpha_1+\alpha_2$. Hence, if such an overestimation control is desired, $\alpha$ should be chosen small. As a default choice we suggest $\alpha = 0.05$. Increasing $\alpha$ yields to a larger detection power (at the price of including more false positives). Hence, one may 'screen' for different $\alpha$ if important events are difficult to detect. The levels $\alpha_1$ and $\alpha_2$ allocates the power between the multiresolution test for detecting events on large scales ($>l_{\max}$) and the local tests to detect events on small scales ($\leq l_{\max}$). We have chosen $\alpha_2=0.04$ and $\alpha_1=0.01$ in our data analysis, since our focus was on detecting short events primarily, while events on larger scales were easier to detect. More weight can be put on $\alpha_1$ if either short events are of less interest or if long events are difficult to detect as well, e.g. since they have a smaller jump size than the short events. The latter is often called \textit{subgating} and was for instance studied in \citep{Hotz.etal.13}. The tuning parameter $l_{\max}$, the largest scale on which local tests are performed to find short events, should be chosen such that all events on larger scales are detected by the previous multiresolution test. This can for instance be determined by Monte-Carlo simulations. In our setting, we choose $l_{\max}=65$, since simulations (not displayed) showed that the multiresolution step in Section \ref{sec:large scales} is able to detect events which contain more than $65$ observations with probability almost one. The correlation matrix is regularized with parameter $\gamma^2=1$, further details can be found in Section III~B in \citep{Peinetal18}. And, as mentioned before, we truncate the kernel and autocorrelation function of the filter at $m=11$ as the autocorrelation function is below $10^{-3}$ afterwards. All of these choices are the default parameters of the function \textit{hilde} and are used in the simulations in Section \ref{sec:simulations} and in the real data application in Section \ref{sec:analysis}.

\section{Simulations}\label{sec:simulations}
In this section we examine the performance of $\HILDE$ in Monte-Carlo simulations. Since to our best knowledge no other model-free method is known that takes into account heterogeneous noise and filtering explicitly, it is difficult to compare $\HILDE$ with other methods. Most similar in spirit are $\JULES$ \citep{Peinetal18}, $\operatorname{HSMUCE}$ and an HMM based approach \citep{Diehn17}. These have been included in a simulation in Section \ref{sec:hmm} for purpose of comparison. The simulation study consists of four parts. First of all, we investigate the detection and idealization of isolated peaks. Secondly, we identify the minimal distance at which $\HILDE$ is able to separate two consecutive peaks. Thirdly, although $\HILDE$ does not rely on a hidden Markov model assumption, we examine its ability to recover the parameters of a Markov model, since a hidden Markov model is a common assumption for ion channel recordings. Finally, we investigate its robustness against violations of the model in Section \ref{sec:model}, in particular against additional $f^2$ and $1/f$ noise.

\subsection{Data generation}\label{sec:datageneration}
We generate all signals and observations accordingly to the heterogeneous ion channel model we described in Section \ref{sec:model} and such that they are in line with the measured data we analyze in Section~\ref{sec:analysis}. This means in particular that amplitudes, dwell times and noise levels of the generated observations are chosen such that they are similar to those of the analyzed datasets. We also simulate a 4-pole Bessel filter with \SI{1}{\kilo\hertz} cut-off frequency and sample the observation at \SI{10}{\kilo\hertz}.\\
The expectation of the observations, given by the convolution of the signal with the truncated kernel $F_m$ of the lowpass Bessel filter, can be computed explicitly. For the errors we oversample by a factor of $100$, i.e., we generate $100$ times as many independent Gaussian observations, discretize the filter accordingly, compute a discrete convolution and rescale the observations such that they have the desired standard deviation.

\subsection{Isolated peak}~\label{sec:singlePeak}
\begin{figure}[htb]
    \centering
    \includegraphics[width=0.4\linewidth]{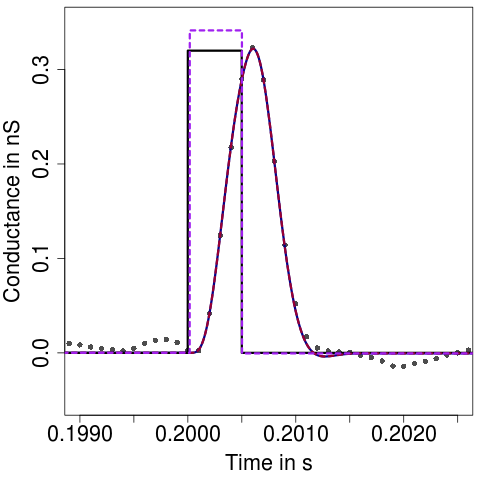} 
\caption{\footnotesize Simulated observations (grey points), true block signal $f$ ({\black \solidrule}) and its convolution ({\black \solidrule}), $\HILDE$s idealization ({\ttlred  \solidrule}) and its convolution ({\myred \solidrule}) with the lowpass 4-pole Bessel filter. $\HILDE$ provides very accurate idealization.}
  \label{fig:singlePeak} 
\end{figure}

In this simulation with $4\,000$ observations we examine the detection and idealization of a single isolated peak. More precisely, in accordance with the model in Section \ref{sec:model} and with the estimated values in Section \ref{sec:analysis} for the observations in Figure \ref{fig:PorBHeteroData}, we choose conductance levels $\valmu_0=\valmu_2=0$, $\valmu_1=0.32$, variances $s_0^2 = s_2^2 = 6.1 \cdot 10^{-5}$ and varying variance $s_1^2\in\{2 \cdot 10^{-4}, 5 \cdot 10^{-4}, 10^{-3}, 2\cdot 10^{-3}, 5\cdot 10^{-3}\}$ to examine the influence of different noise levels. Note that $s_1^2=10^{-3}$ is roughly the noise level in the measurements in Section \ref{sec:analysis}. Moreover, we simulate changes at $\tau_1=2000 / \sr$ and $\tau_2 = (2000 + \ell) / \sr$, c.f.~\eqref{eq:signal}, and are interested in how well $\HILDE$ detects the peak and idealizes the locations $\tau_1$ and $\tau_2$ and the level $l_1$ as a function of $\ell$, the length (relative to the sampling rate $\sr$) of the peak. For $\ell=5$, Figure~\ref{fig:singlePeak} shows an example of the simulated data as well as the idealizations by $\HILDE$ and their convolutions with the Bessel filter in a neighborhood of the peak. Tables~\ref{tab_singlePeak_hetero_detection}-\ref{tab_singlePeak_hetero_estimationLevel} summarize our results based on $10\,000$ repetitions for $\ell=2,3,5$.\\
To this end, we count how often the signal is \textit{correctly identified}, i.e., only the peak and no other change is detected. More precisely, we define the peak as \textit{detected} if there exists a $j$ such that $|\hat{\tau}_j - \tau_1| < m / \sr$ and $|\hat{\tau}_{j+1} - \tau_2| < m / \sr$ as a peak is shifted at most $m / \sr$ by the filter. If only one change but not a peak is within these boundaries we do not count it as a true detection, but also not as a \textit{false positive}, whereas all other changes are counted as \textit{false positives}. For the estimated locations and the level we only consider cases where the peak is detected and report the mean square error, the bias and the standard deviation.

\begin{table}[!htb]
\centering
 \caption{\footnotesize Performance of $\HILDE$ in idealizing a signal with an isolated peak in different settings. They differ in the amount of open channel noise and the length of the peak. More precisely, it has changes at $\tau_1=0.2$ and $\tau_2 = \tau_1 + \ell / \sr$, $\ell = 2,3,5$, conductance levels $\valmu_0=\valmu_2=0$, $\valmu_1=0.32$, variances $s_0^2 = s_2^2 = 6.1 \cdot 10^{-5}$ and varying variance $s_1^2\in\{2 \cdot 10^{-4}, 5 \cdot 10^{-4}, 10^{-3}, 2\cdot 10^{-3}, 5\cdot 10^{-3}\}$. Results are based on $10\,000$ pseudo samples. An example, $s_1^2 = 10^{-3}$ and $\ell=5$, is given in Figure~\ref{fig:singlePeak}.}~\label{tab_singlePeak_hetero_detection}
\scalebox{0.75}{
  \begin{tabular}{l N N N N}
  \hline
  Setting & Length ($\ell$) & Correctly identified ($\%$) & Detected ($\%$) & False positive (Mean)\\
  \hline
$s_1^2 = 2\cdot 10^{-4}$ & 2 & 99.96 & 100.00 & 0.0008 \\ 
$s_1^2 = 5\cdot 10^{-4}$ & 2 & 99.96 & 100.00 & 0.0008 \\ 
$s_1^2 = 10^{-3}$ & 2 & 99.94 & 99.98 & 0.0010 \\ 
$s_1^2 = 2\cdot 10^{-3}$ & 2 & 99.07 & 99.11 & 0.0014 \\ 
$s_1^2 = 5\cdot 10^{-3}$ & 2 & 90.04 & 90.08 & 0.0042 \\ 
    \hline  
$s_1^2 = 2\cdot 10^{-4}$ & 3 & 99.97 & 100.00 & 0.0006 \\ 
$s_1^2 = 5\cdot 10^{-4}$ & 3 & 99.97 & 100.00 & 0.0006 \\ 
$s_1^2 = 10^{-3}$ & 3 & 99.97 & 100.00 & 0.0006 \\ 
$s_1^2 = 2\cdot 10^{-3}$ & 3 & 99.93 & 99.96 & 0.0006 \\ 
$s_1^2 = 5\cdot 10^{-3}$ & 3 & 96.08 & 96.11 & 0.0024 \\ 
    \hline 
$s_1^2 = 2\cdot 10^{-4}$ & 5 & 99.95 & 100.00 & 0.0010 \\ 
$s_1^2 = 5\cdot 10^{-4}$ & 5 & 99.95 & 100.00 & 0.0010 \\ 
$s_1^2 = 10^{-3}$ & 5 & 99.95 & 100.00 & 0.0010 \\ 
$s_1^2 = 2\cdot 10^{-3}$ & 5 & 99.94 & 100.00 & 0.0012 \\ 
$s_1^2 = 5\cdot 10^{-3}$ & 5 & 99.42 & 99.48 & 0.0018 \\ 
  \hline
  \end{tabular}
}
\end{table}

\begin{table}[!hb]
\centering
 \caption{\footnotesize Performance of $\HILDE$ in idealizing a signal with an isolated peak in different settings. They differ in the amount of open channel noise and the length of the peak. More precisely, it has changes at $\tau_1=0.2$ and $\tau_2 = \tau_1 + \ell / \sr$, $\ell = 2,3,5$, conductance levels $\valmu_0=\valmu_2=0$, $\valmu_1=0.32$, variances $s_0^2 = s_2^2 = 6.1 \cdot 10^{-5}$ and varying variance $s_1^2\in\{2 \cdot 10^{-4}, 5 \cdot 10^{-4}, 10^{-3}, 2\cdot 10^{-3}, 5\cdot 10^{-3}\}$. Results are based on $10\,000$ pseudo samples and are given as multiples of the sampling rate $\sr = 10^4$. An example, $s_1^2 = 10^{-3}$ and $\ell=5$, is given in Figure~\ref{fig:singlePeak}.}\label{tab_singlePeak_hetero_estimationLocation}
\scalebox{0.75}{
  \begin{tabular}{l N N N N N N N N}
  \hline
   Setting & Length ($\ell$) 
  & $\sr^2\MSE(\hat{\tau}_1)$ & $\sr\BIAS(\hat{\tau}_1)$ & $\sr\SD(\hat{\tau}_1)$ 
  & $\sr^2\MSE(\hat{\tau}_2)$ & $\sr\BIAS(\hat{\tau}_2)$ & $\sr\SD(\hat{\tau}_2)$\\ 
  \hline
$s_1^2 = 2\cdot 10^{-4}$ & 2 & 0.0331 & 0.0092 & 0.1818 & 0.0381 & -0.0076 & 0.1951 \\ 
$s_1^2 = 5\cdot 10^{-4}$ & 2 & 0.0515 & 0.0113 & 0.2267 & 0.0427 & -0.0115 & 0.2062 \\ 
$s_1^2 = 10^{-3}$ & 2 & 0.0677 & 0.0255 & 0.2590 & 0.0595 & -0.0266 & 0.2424 \\ 
$s_1^2 = 2\cdot 10^{-3}$ & 2 & 0.1532 & 0.0935 & 0.3801 & 0.1570 & -0.0942 & 0.3848 \\ 
$s_1^2 = 5\cdot 10^{-3}$ & 2 & 0.6628 & 0.3275 & 0.7454 & 0.6223 & -0.3252 & 0.7188 \\ 
    \hline
$s_1^2 = 2\cdot 10^{-4}$ & 3 & 0.0120 & 0.0001 & 0.1097 & 0.0117 & 0.0010 & 0.1083 \\ 
$s_1^2 = 5\cdot 10^{-4}$ & 3 & 0.0177 & 0.0040 & 0.1329 & 0.0177 & -0.0022 & 0.1332 \\ 
$s_1^2 = 10^{-3}$ & 3 & 0.0391 & 0.0181 & 0.1970 & 0.0388 & -0.0152 & 0.1965 \\ 
$s_1^2 = 2\cdot 10^{-3}$ & 3 & 0.1569 & 0.0846 & 0.3870 & 0.1533 & -0.0802 & 0.3833 \\ 
$s_1^2 = 5\cdot 10^{-3}$ & 3 & 1.1681 & 0.4668 & 0.9748 & 1.1058 & -0.4515 & 0.9498 \\ 
    \hline 
$s_1^2 = 2\cdot 10^{-4}$ & 5 & 0.0070 & -0.0014 & 0.0835 & 0.0084 & 0.0013 & 0.0914 \\ 
$s_1^2 = 5\cdot 10^{-4}$ & 5 & 0.0176 & 0.0036 & 0.1326 & 0.0189 & -0.0037 & 0.1375 \\ 
$s_1^2 = 10^{-3}$ & 5 & 0.0572 & 0.0217 & 0.2381 & 0.0591 & -0.0215 & 0.2421 \\ 
$s_1^2 = 2\cdot 10^{-3}$ & 5 & 0.2473 & 0.0985 & 0.4874 & 0.2464 & -0.0976 & 0.4867 \\ 
$s_1^2 = 5\cdot 10^{-3}$ & 5 & 2.0808 & 0.6363 & 1.2946 & 2.1605 & -0.6520 & 1.3174 \\   
  \hline
  \end{tabular}
}
\end{table}

\begin{table}[!htb]
\centering
 \caption{\footnotesize Performance of $\HILDE$ in idealizing a signal with an isolated peak in different settings. They differ in the amount of open channel noise and the length of the peak. More precisely, it has changes at $\tau_1=0.2$ and $\tau_2 = \tau_1 + \ell / \sr$, $\ell = 2,3,5$, conductance levels $\valmu_0=\valmu_2=0$, $\valmu_1=0.32$, variances $s_0^2 = s_2^2 = 6.1 \cdot 10^{-5}$ and varying variance $s_1^2\in\{2 \cdot 10^{-4}, 5 \cdot 10^{-4}, 10^{-3}, 2\cdot 10^{-3}, 5\cdot 10^{-3}\}$. Results are based on $10\,000$ pseudo samples. An example, $s_1^2 = 10^{-3}$ and $\ell=5$, is given in Figure~\ref{fig:singlePeak}.}\label{tab_singlePeak_hetero_estimationLevel}
\scalebox{0.75}{
  \begin{tabular}{l N N N N N N N N}
  \hline 
    Setting & Length ($\ell$) 
  & $\MSE(\hat{\valmu}_1)$ & $\BIAS(\hat{\valmu}_1)$ & $\SD(\hat{\valmu}_1)$\\ 
  \hline
$s_1^2 = 2\cdot 10^{-4}$ & 2 & 0.1320 & 0.0194 & 0.3628 \\ 
$s_1^2 = 5\cdot 10^{-4}$ & 2 & 0.6953 & 0.0322 & 0.8333 \\ 
$s_1^2 = 10^{-3}$ & 2 & 2.3290 & 0.0888 & 1.5236 \\ 
$s_1^2 = 2\cdot 10^{-3}$ & 2 & 21.7896 & 0.7801 & 4.6025  \\ 
$s_1^2 = 5\cdot 10^{-3}$ & 2 & 294.6030 & 5.5640 & 16.2380 \\ 
    \hline
$s_1^2 = 2\cdot 10^{-4}$ & 3 & 0.0002 & 0.0009 & 0.0152 \\ 
$s_1^2 = 5\cdot 10^{-4}$ & 3 & 0.0007 & 0.0018 & 0.0259\\ 
$s_1^2 = 10^{-3}$ & 3 & 0.0023 & 0.0055 & 0.0473\\ 
$s_1^2 = 2\cdot 10^{-3}$ & 3 & 1.7338 & 0.0806 & 1.3143 \\ 
$s_1^2 = 5\cdot 10^{-3}$ & 3 & 334.7102 & 5.0891 & 17.5739  \\ 
    \hline 
$s_1^2 = 2\cdot 10^{-4}$ & 5 & 0.0001 & 0.0001 & 0.0077 \\ 
$s_1^2 = 5\cdot 10^{-4}$ & 5 & 0.0003 & 0.0007 & 0.0179 \\ 
$s_1^2 = 10^{-3}$ & 5 & 0.0013 & 0.0033 & 0.0354 \\ 
$s_1^2 = 2\cdot 10^{-3}$ & 5 & 0.0055 & 0.0158 & 0.0725 \\ 
$s_1^2 = 5\cdot 10^{-3}$ & 5 & 203.0129 & 2.7499 & 13.9811 \\ 
  \hline
  \end{tabular}
}
\label{tab:valueheterogeneous}
\end{table}

In most scenarios, $\HILDE$ has a good detection power and detects almost no false positives, see Table \ref{tab_singlePeak_hetero_detection}. Only for a five times larger variance than in the real data and when $l\leq 3$ few events are missed. In Tables \ref{tab_singlePeak_hetero_estimationLocation} and \ref{tab_singlePeak_hetero_estimationLevel} we found that idealization of the locations $\tau_1$ and $\tau_2$ and the conductance value $\valmu_1$ works well for variances similar to the real data, but has some issues when the variance of the peak is larger, in particular in the scenario of a five times larger variance. For such observations it might be desirable to take into account the heterogeneous noise in the deconvolution step, see Section \ref{sec:discussionalternatives} for more details. For smaller variances the results for estimating the locations are better when the peak is longer, but for larger variances results are even worse when the peak is longer. An explanation might be two effects with opposite influences. The conductance change provide more information when the peak is longer, but then also the overall variance of the observations is larger which reduces estimation accuracy. Estimation of the level $\valmu_1$ is always more accurate when the peak is longer. It seems that here the first effect dominates.\\ 
All in all, these simulations confirm that $\HILDE$ performs very well for observations comparable to them in Section \ref{sec:analysis}.

\subsection{Separation of two consecutive peaks}~\label{sec:consecutivePeaks}
\begin{figure}[htb]
  \begin{subfigure}[t]{0.32\linewidth}
    \centering
    \includegraphics[width=\linewidth]{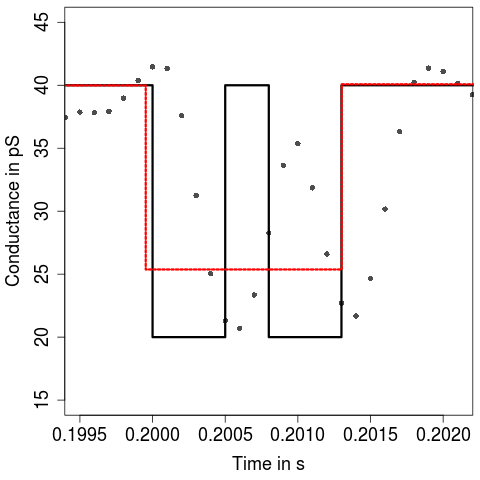} 
    \caption{\footnotesize{No separation in the detection step, $d = 3$}}
    \label{fig:twoPeaks_b} 
  \end{subfigure}
  \begin{subfigure}[t]{0.32\linewidth}
    \centering
    \includegraphics[width=\linewidth]{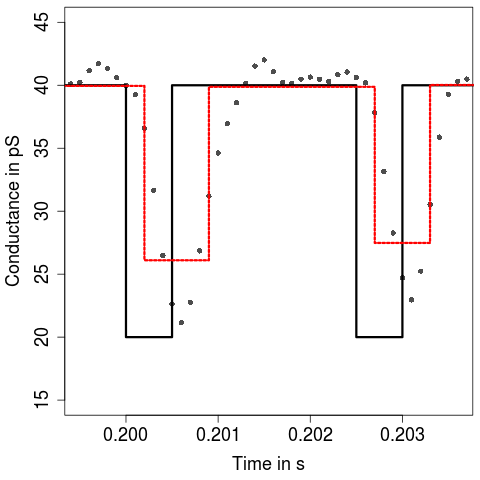} 
    \caption{\footnotesize{No separation in the deconvolution, $d = 20$}}
    \label{fig:twoPeaks_c} 
  \end{subfigure} 
  \begin{subfigure}[t]{0.32\linewidth}
    \centering
    \includegraphics[width=\linewidth]{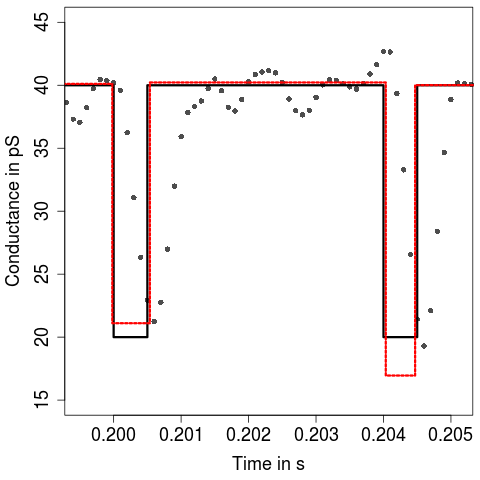} 
    \caption{\footnotesize{Perfect separation, $d = 35$}}
    \label{fig:twoPeaks_d} 
  \end{subfigure}
  \caption{\footnotesize{Data $y_i = F \ast (f(i/n) + \sigma_0 \epsilon_i)$ (grey points), where $\sigma_0=1.4$, $\epsilon_i$ is gaussian white noise and the signal $f$  has two consecutive peaks comprised of the levels $l_0{\black = l_2 = l_4} = 40$, $l_1{\black = l_3} = 20$ and change-points $\tau_1 = 2000 / \sr$, $\tau_2 = \tau_1 + 5/\sr$, $\tau_3 = \tau_2 + d$ and $\tau_4 = \tau_3 + 5/\sr$. True signal ({\black \solidrule}) and $\JULES$ idealization ({\ttlred  \solidrule}). Idealization coincides with reconstruction from the detection step if separation fails in the deconvolution step.}}
  \label{fig:twoPeaks} 
\end{figure}

To examine how well $\HILDE$ separates two consecutive peaks we perform the same simulations as in Section 4.3 in \citep{Peinetal18}, since results are identical for homogeneous and heterogeneous noise as separation depends on the method and distance between the peaks but not on the noise level. More precisely, we consider a signal $f$ with changes at $\tau_1 = 2\,000 / \sr$, $\tau_2 = \tau_1 + 5/\sr$, $\tau_3 = \tau_2 + d$ and $\tau_4 = \tau_3 + 5/\sr$, , with $\tau_0 = 0$ and $\tau_{\operatorname{end}} = 4\,000/\sr$ and levels $l_0=l_2=l_4=\SI{40}{\pico\siemens}$ and $l_1=l_3=\SI{20}{\pico\siemens}$. Hence $d$ is the distance between the two peaks. We distinguish between perfect separation, i.e., the detection step of $\HILDE$ identifies the two peaks (4 changes) and the local deconvolution yields idealizations for the four levels (illustrated in Figure \ref{fig:twoPeaks_d}). Secondly, separation fails in the detection step, i.e., the multiresolution reconstruction recognizes only 2 changes and identifies one peak whose level can be further deconvolved (illustrated in Figure \ref{fig:twoPeaks_b}). Finally, separation fails in the deconvolution step, i.e., $\HILDE$ identifies two peaks but the distance is so small that the deconvolution step cannot separate them, in other words, no long segment is in between (illustrated in Figure \ref{fig:twoPeaks_c}).

\begin{figure}[!hb]
  \centering
    \includegraphics[width=.98\textwidth, keepaspectratio]{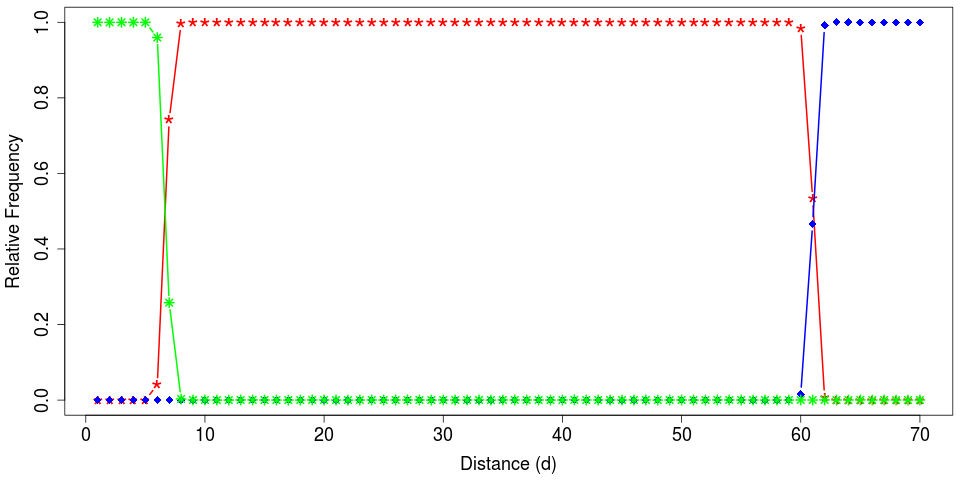}
    \caption{\footnotesize{Results for $\HILDE$ assuming heterogeneous noise in idealizing two consecutive peaks separated by distance $d$. Its frequencies for no separation in the detection step (green), for successful detection, but no separation in the deconvolution step (red) and for successful detection and deconvolution (blue). Results are based on $10\,000$ simulations for each value of $d$.}}
    \label{fig:separation} 
\end{figure}

Figure~\ref{fig:separation} shows the frequency at which each scenario occurred as a function of $d$, the distance between the two peaks, in $10\,000$ simulations for each value of $d = \{1, 2, \ldots, 70\}$. We found that the two peaks are detected if $d > 6$, but separation in the detection step and hence an appropriate idealization requires $d \geq 62$. Hence, in Section \ref{sec:analysis} events have to be separated by more than $\SI{6.2}{\milli\second}$ to be idealized appropriately. In comparison, we found that events are on average separated by $\SI[parse-numbers = false]{1 / (2.67 + 4.50)}{\second}\approx \SI{0.14}{\second}$ which shows that this limitation is not an issue for the analyzed PorB recordings.

\subsection{Hidden Markov model}~\label{sec:hmm}
\begin{figure}[!htb]
\centering
\begin{subfigure}{0.32\textwidth}
\includegraphics[width = \textwidth]{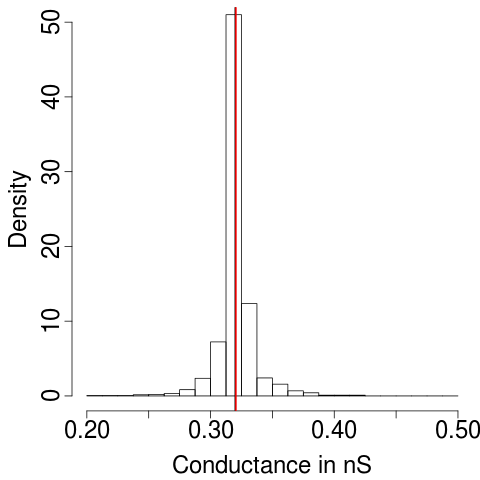}
\subcaption{$\HILDE$}
\label{subfig:A:amplitudeHILDE}
\end{subfigure}
\begin{subfigure}{0.32\textwidth}
\includegraphics[width = \textwidth]{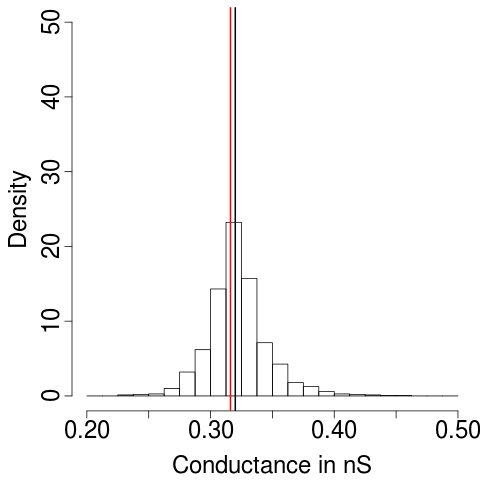}
\subcaption{$\JULES$}
\label{subfig:A:amplitudeJULES}
\end{subfigure}
\begin{subfigure}{0.32\textwidth}
\includegraphics[width = \textwidth]{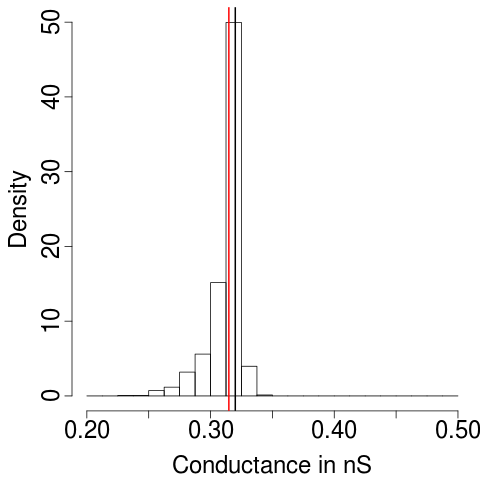}
\subcaption{$\HSMUCE$}
\label{subfig:A:amplitudeHSMUCE}
\end{subfigure}
\caption{Histograms of the estimated amplitudes using various idealization approaches. The black line (for $\HILDE$ hidden by the red line) indicates the true amplitude of $0.32$ and the red line the estimated amplitudes of $0.3205$, $0.3161$ and $0.3148$ by the half sample mode using the idealizations from $\HILDE$, $\JULES$ and $\HSMUCE$, respectively.}
\label{fig:HMMamplitude}
\end{figure}

In this section we simulate data from a three state hidden Markov model. Since hidden Markov models are often assumed for ion channel recordings, it is instructive to investigate the methods in such a scenario. We simulate observations that resemble the PorB data we analyze in Section \ref{sec:analysis}. More precisely, we have expectations $\SI{0}{\nano\siemens}$, $\SI{0}{\nano\siemens}$ and $\SI{0.32}{\nano\siemens}$ as well as standard deviations $\SI{0.0078}{\nano\siemens}$, $\SI{0.0078}{\nano\siemens}$ and $\SI{0.0316}{\nano\siemens}$, i.e. the variances are $\SI[parse-numbers = false]{6.1\cdot 10^{-5}}{(\nano\siemens)^2}$ and $\SI[parse-numbers = false]{10^{-3}}{(\nano\siemens)^2}$. The dwell times in the first, second and third state are exponentially distributed with rates $\SI{20}{\hertz}$, $\SI{400}{\hertz}$ and $\SI{7}{\hertz}$, respectively. The process always jumps from the first or second state to the third state, i.e., no transitions between the first and second state are allowed. And it jumps from the third state with probability $2/3$ to the first state and with probability $1/3$ to the second state. We generate five time series with $600\,000$ observations, each. Each trace looks similar to the observations in Figure \ref{fig:PorBHeteroData} and hence we refrain from showing an example.\\
We analyze these data sets with $\HILDE$ and for purpose of comparison with $\JULES$ \citep{Peinetal18}, $\operatorname{HSMUCE}$ \citep{Pein.Sieling.Munk.16} and an HMM based approach which assumes the true three state model, i.e. three states, whereby two have the same expectations and variances and no transitions are allowed between them. We used
\[\mu = (0, 0, 0.32)^T,\ s = (0.0102, 0.0102, 0.0321)^T,\ P = \left( \begin{matrix} 0.999 & 0 & 0.001\\ 
0 & 0.99 & 0.01\\ 0.02 & 0.03 & 0.95 \end{matrix} \right).\]  
as starting values for the Baum-Welch algorithm. Those standard deviations were determined by taking the empirical standard deviation of all observations below and above $0.2$, respectively. Idealizations are obtained by using a Viterbi algorithm.\\
The Baum-Welch algorithm estimated the following parameters
\[\mu = (0, 0, 0.3177)^T,\ s = (0.0079, 0.0079, 0.0374)^T,\ P = \left( \begin{matrix} 0.999356 & 0 & 0.000644\\ 
0 & 0.997329 & 0.002671\\ 0.002617 & 0.000089 &  0.997294 \end{matrix} \right).\]
We will discuss the estimated transition matrix later in comparison with the other approaches and when we also discuss the results using the Viterbi algorithm. The estimated expectations and standard deviations are accurate. For the other approaches we show in Figure \ref{fig:HMMamplitude} histograms of the estimated amplitudes of all events with an amplitude between $\SI{0.2}{\nano\siemens}$ and $\SI{0.5}{\nano\siemens}$. 

\begin{figure}[!htb]
\centering
\begin{subfigure}{0.24\textwidth}
\includegraphics[width = \textwidth]{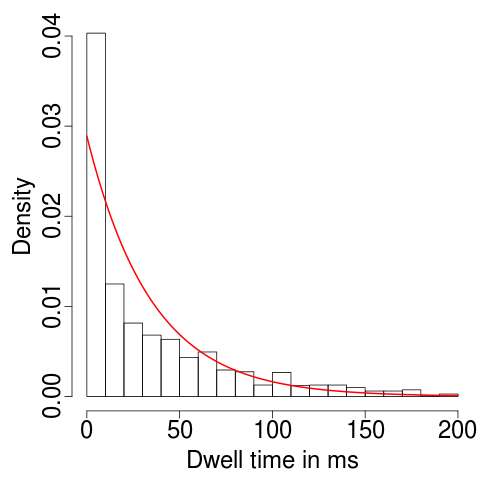}
\subcaption{$\HILDE$}
\label{subfig:HMMdwellHILDE}
\end{subfigure}
\begin{subfigure}{0.24\textwidth}
\includegraphics[width = \textwidth]{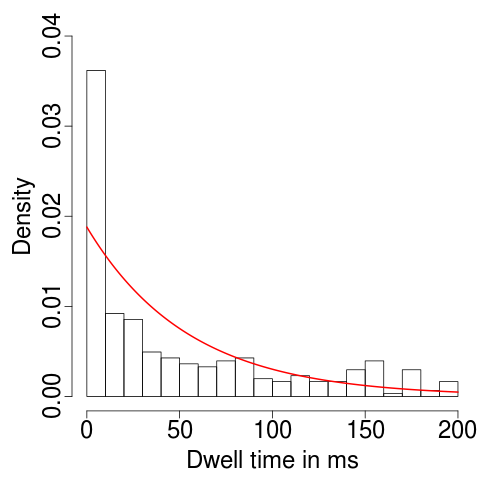}
\subcaption{$\JULES$}
\label{subfig:HMMdwellJULES}
\end{subfigure}
\begin{subfigure}{0.24\textwidth}
\includegraphics[width = \textwidth]{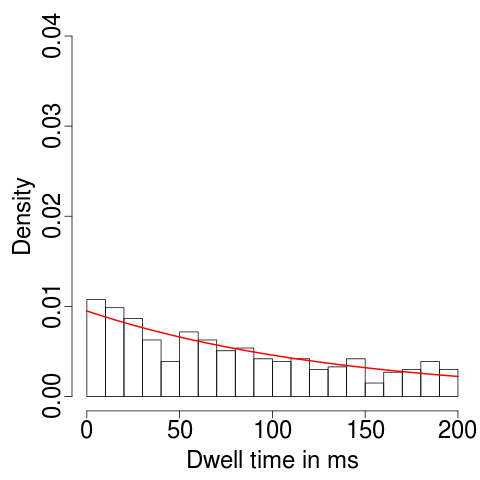}
\subcaption{$\HSMUCE$}
\label{subfig:HMMdwellHSMUCE}
\end{subfigure}
\begin{subfigure}{0.24\textwidth}
\includegraphics[width = \textwidth]{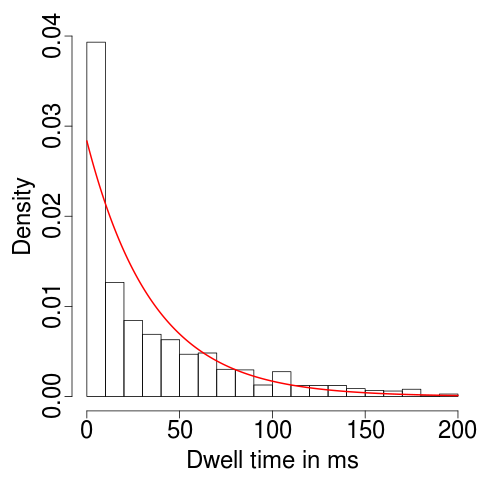}
\subcaption{$\HMM$}
\label{subfig:HMMdwellHMM}
\end{subfigure}
\caption{Histograms of the dwell times in the closed state with exponential fits (red). We rescaled all lines such that the area under them are standardized to one to make them comparable to the histograms.}
\label{fig:HMMdwell}
\end{figure}

We found in Figure \ref{fig:HMMamplitude} that all approaches estimate the amplitude accurately. The estimated amplitudes of $\HSMUCE$ are skewed, but the final estimation is still decent.\\
We continue with an analysis of the dwell times. To this end, we consider from now on all events with estimated conductance level between $\SI{-0.05}{\nano\siemens}$ and $\SI{0.05}{\nano\siemens}$ as a closed event and between $\SI{0.15}{\nano\siemens}$ and $\SI{2}{\nano\siemens}$ as an open event, while all other events are considered as artifacts and are ignored. Figure \ref{fig:HMMdwell} shows histograms of the dwell times in the closed state for various approaches.

\begin{figure}[!htb]
\centering
\begin{subfigure}{0.24\textwidth}
\includegraphics[width = \textwidth]{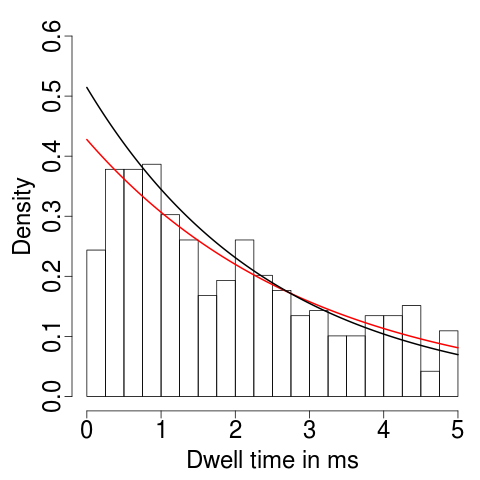}
\subcaption{$\HILDE$}
\label{subfig:HMMdwellShortHILDE}
\end{subfigure}
\begin{subfigure}{0.24\textwidth}
\includegraphics[width = \textwidth]{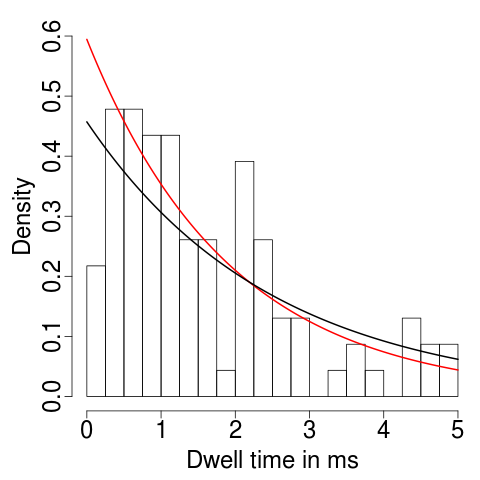}
\subcaption{$\JULES$}
\label{subfig:HMMdwellShortJULES}
\end{subfigure}
\begin{subfigure}{0.24\textwidth}
\includegraphics[width = \textwidth]{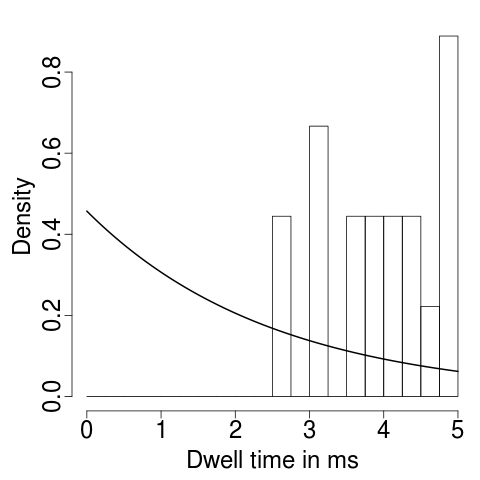}
\subcaption{$\HSMUCE$}
\label{subfig:HMMdwellShortHSMUCE}
\end{subfigure}
\begin{subfigure}{0.24\textwidth}
\includegraphics[width = \textwidth]{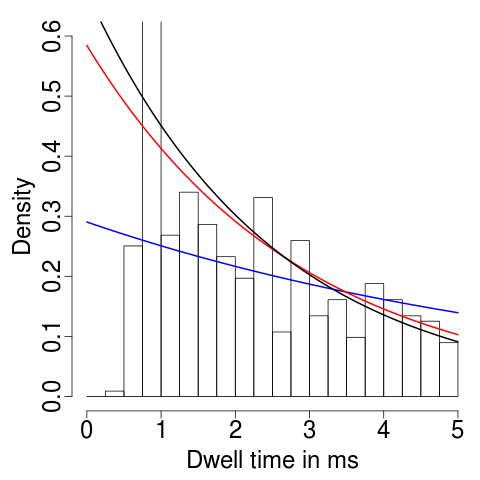}
\subcaption{$\HMM$}
\label{subfig:HMMdwellShortHMM}
\end{subfigure}
\caption{Histograms of the dwell times in the closed state with a length between $\SI{0.1}{\milli\second}$ and $\SI{5}{\milli\second}$ to analyze short events, together with the true exponential distribution (black line) and exponential fits (red) that are corrected for missed events. We rescaled all lines such that the area under them are standardized to one to make them comparable to the histograms. This explains why the true distribution does not look always the same. The blue line in the $\HMM$ plot indicates an exponential fit when the same missed event correction is used that is applied to the results obtained by $\HILDE$.}
\label{fig:HMMdwellShort}
\end{figure}

\begin{figure}[!htb]
\centering
\begin{subfigure}{0.24\textwidth}
\includegraphics[width = \textwidth]{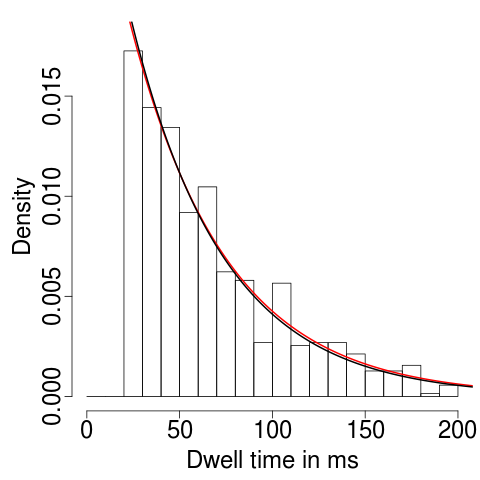}
\subcaption{$\HILDE$}
\label{subfig:HMMdwellLongHILDE}
\end{subfigure}
\begin{subfigure}{0.24\textwidth}
\includegraphics[width = \textwidth]{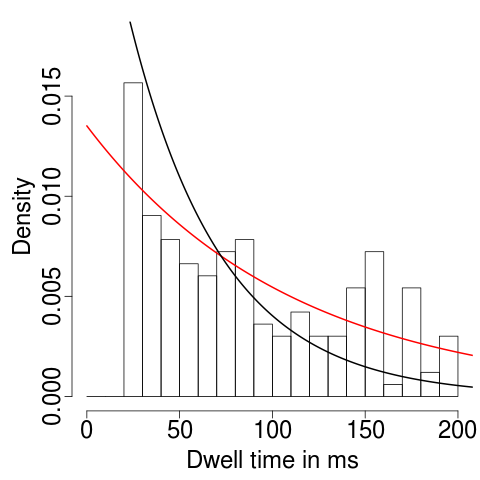}
\subcaption{$\JULES$}
\label{subfig:HMMdwellLongJULES}
\end{subfigure}
\begin{subfigure}{0.24\textwidth}
\includegraphics[width = \textwidth]{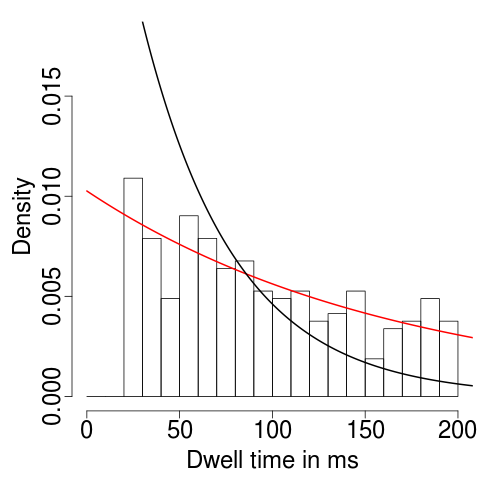}
\subcaption{$\HSMUCE$}
\label{subfig:HMMdwellLongHSMUCE}
\end{subfigure}
\begin{subfigure}{0.24\textwidth}
\includegraphics[width = \textwidth]{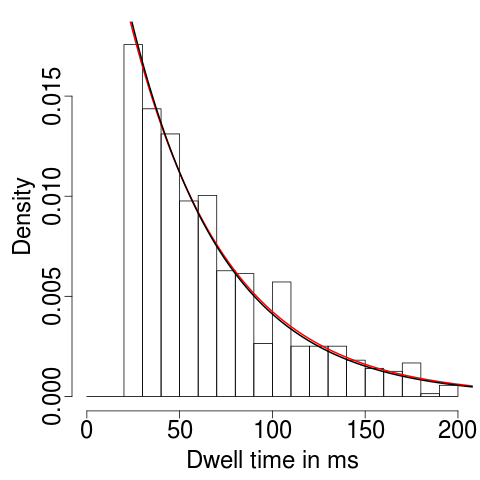}
\subcaption{$\HMM$}
\label{subfig:HMMdwellLongHMM}
\end{subfigure}
\caption{Histograms of the dwell times in the closed state with a length between $\SI{20}{\milli\second}$ and $\SI{200}{\milli\second}$ to analyze long events, together with the true exponential distribution (black line) and exponential fits (red) that are corrected for missed events. We rescaled all lines such that the area under them are standardized to one to make them comparable to the histograms. This explains why the true distribution does not look always the same.}
\label{fig:HMMdwellLong}
\end{figure}

We see that with the exception of $\HSMUCE$ (it misses short events) none on the histograms look exponentially distributed, since we have a mixture of short and long events. Hence, in Figures \ref{fig:HMMdwellShort} and \ref{fig:HMMdwellLong} we will analyze short and long events separately. To this end, we say an event is short if its dwell time is between $\SI{0.1}{\milli\second}$ and $\SI{5}{\milli\second}$ and long if its dwell time is between $\SI{20}{\milli\second}$ and $\SI{200}{\milli\second}$. To estimate the rates, we apply a missed event correction like in \citep{Peinetal18}.

We found from Figures \ref{fig:HMMdwellShort} and \ref{fig:HMMdwellLong} that $\HILDE$ recovers in both cases the exponential distribution very well and estimates both rates of \SI{400}{\hertz} and \SI{20}{\hertz} with \SI{332.59}{\hertz} and \SI{19.2629}{\hertz} accurately. In comparison, $\JULES$ is not able to deconvolve all events due to the detection of additional spurious events, compare Figure \ref{fig:PorBHeteroJULES}. The rate for the short events is with \SI{520.16}{\hertz} still accurately, but the rate for the long events is with \SI{9.0682}{\hertz} significantly underestimated. Notably the dwell times are still (almost) exponentially distributed. $\HSMUCE$ misses short events, in total it has detected only $18$ short events. Hence, a rate for the short events cannot be estimated. The rate for the long events is with \SI{6.0182}{\hertz} underestimated as well. The hidden Markov approach estimated with \SI{347.44}{\hertz} and \SI{19.3949}{\hertz} both rates accurately. However, since this approach misses very short events, for the rate for the short events we had to apply a stricter missed event correction that takes into account only events with a length of at least $\SI{0.75}{\milli\second}$. Hence, at least in the used form the hidden Markov approach is less favorable to analyze short events (since its corrected estimate is based on less event and hence will have a larger variance). This is remarkably, since the idealization on very short temporal scales is considered to be a strength of hidden Markov approaches. Finally, the estimated exit probabilities by the Baum-Welch algorithmus of $0.002671$ and $0.000644$ corresponds to estimated rates of \SI{26.71}{\hertz} and \SI{6.44}{\hertz} which is much worse than the rates estimated using the idealizations obtained by the Viterbi algorithm.\\
We are now analyzing how often closing events occur. To this end, we analyze the dwell times in the open state or in other words the distance between two closing events. Moreover, we analyze the proportions of short and long events. Therefore, we divide the number of detected events by the estimated probability that such an event is detected assuming an exponential distribution for the dwell times.

\begin{figure}[!htb]
\centering
\begin{subfigure}{0.24\textwidth}
\includegraphics[width = \textwidth]{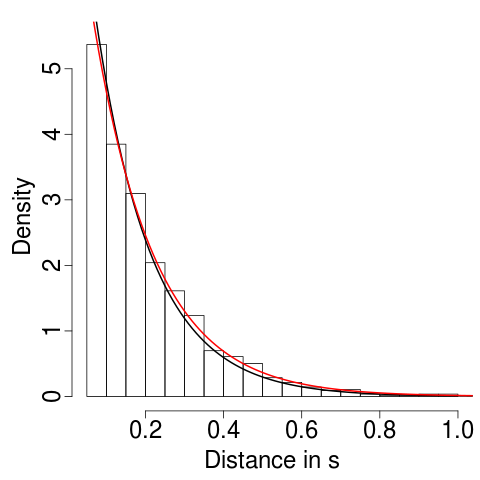}
\subcaption{$\HILDE$}
\label{subfig:HMMdistanceHILDE}
\end{subfigure}
\begin{subfigure}{0.24\textwidth}
\includegraphics[width = \textwidth]{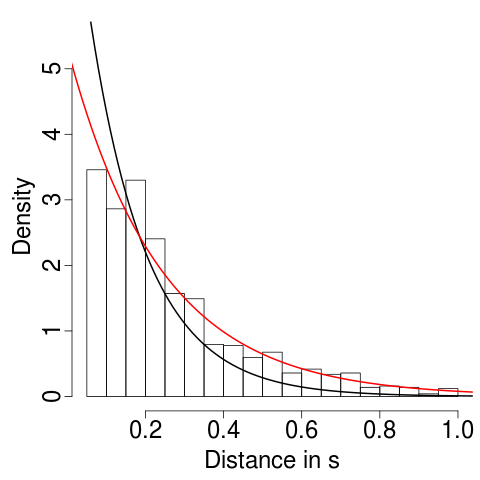}
\subcaption{$\JULES$}
\label{subfig:HMMdistanceJULES}
\end{subfigure}
\begin{subfigure}{0.24\textwidth}
\includegraphics[width = \textwidth]{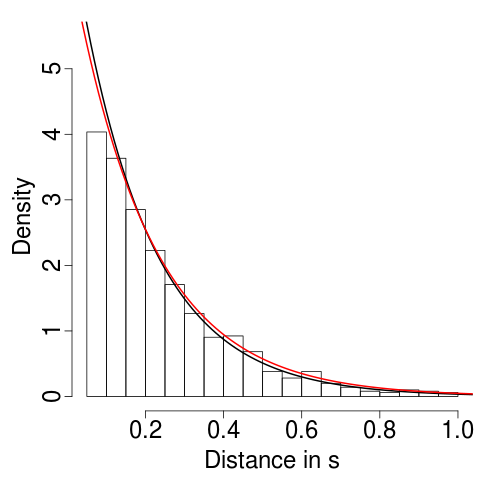}
\subcaption{$\HSMUCE$}
\label{subfig:HMMdistanceHSMUCE}
\end{subfigure}
\begin{subfigure}{0.24\textwidth}
\includegraphics[width = \textwidth]{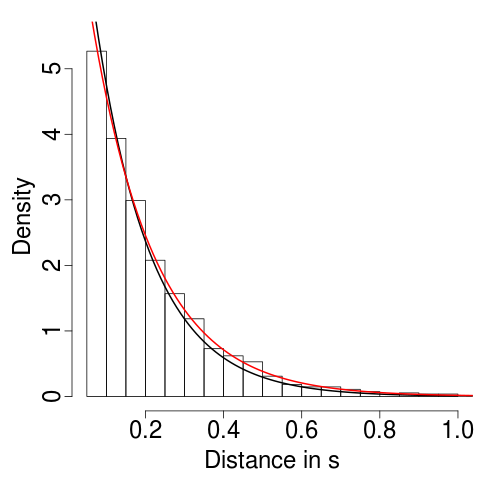}
\subcaption{$\HMM$}
\label{subfig:HMMdistanceHMM}
\end{subfigure}
\caption{Histograms of the dwell times in the open state, together with the true exponential distribution (black line) and exponential fits (red) that are corrected for missed events. We rescaled all lines such that the area under them are standardized to one to make them comparable to the histograms. This explains why the true distribution does not look always the same.}
\label{fig:HMMdistance}
\end{figure}

We found from Figure \ref{fig:HMMdistance} that $\HILDE$, $\HSMUCE$ and the hidden Markov approach recover the exponential distribution very well and estimate the rate of \SI{7}{\hertz} with \SI{6.4007}{\hertz}, \SI{6.4919}{\hertz} and \SI{6.2389}{\hertz} accurately. Only $\JULES$ underestimates the rate because of previously explained reasons with \SI{4.3362}{\hertz} a bit. $\HILDE$, $\JULES$ and $\HMM$ estimated with $0.3608$, $0.2982$ and $0.4082$, respectively, the proportion of short events decently, recall that the truth is $1/3$. This number could not be determined using $\HSMUCE$, since it misses almost all short events. Once again, the Baum-Welch provides with \SI{27.06}{\hertz} and $0.0329$ much worse results.\\
All in all, we found that $\HILDE$ was indeed able to recover all parameters very well. All other model-free idealization methods had at least one massive problem. The hidden Markov approach might be usable, but requires a more restrictive missed event correction and is also more complicated to apply. One should also keep in mind that we used the true parametric model class as prior knowledge.

\subsection{Robustness}\label{sec:robustness}
The model we proposed in Section \ref{sec:model} is a good assumption for ion channel recording at presence of open channel noise. However, in some patch clamp recordings additional high frequency $f^2$ (violet) and long tailed $1/f$ (pink) noise components have been observed, for a more detailed discussion see \citep{Neher.Sakmann.76, venkataramanan1998identification, levis1993use} and the references therein. Thereto, in this section we examine how robust $\HILDE$ is against such noise components. To this end, we revisit the simulation setting from Section \ref{sec:singlePeak} with $s_1 = 10^{-3}$ only.\\
For the violet noise we use as suggested by \citep{venkataramanan1998} a moving average process with coefficients $0.8$ and $-0.6$. For the pink noise we use the algorithm available on \textit{https://github.com/Stenzel/newshadeofpink}. We assume that the pink noise is globally present. More precisely, we reduce the previously present noise by a factor of $1/2$ and add pink noise which is scaled such that its standard deviation is equal to $1/2 \sqrt{6.1 \cdot 10^{-5}}$ (half of the standard deviation in the background in Section \ref{sec:singlePeak}). For the high frequency violet noise we consider the setting that the new noise component is state-dependent as well. In other words, we generated errors from such a moving average process and convolved them with the kernel of the lowpass filter instead of assuming white noise errors.

\begin{table}[ht]
\centering
 \caption{\footnotesize Robustness of $\HILDE$ against additional noise components in idealizing a signal with an isolated peak having changes at $\tau_1=0.2$ and $\tau_2 = \tau_1 + \ell / \sr$, $\ell = 2,3,5$, conductance levels $\valmu_0=\valmu_2=0$, $\valmu_1=0.32$, variances $s_0^2 = s_2^2 = 6.1 \cdot 10^{-5}$ and $s_1^2=10^{-3}$. Results are based on $10\,000$ pseudo samples.}~\label{tab_singlePeak_robustness}
\scalebox{0.75}{
  \begin{tabular}{l N N N N}
  \toprule[1.25pt]  
  Noise type & Length ($\ell$) & Correctly identified ($\%$) & Detected ($\%$) & False positive (Mean)\\
  \hline
  \\
  White noise & 2 & 99.94 & 99.98 & 0.0010 \\ 
  $f^2$ noise & 2 & 99.94 & 99.98 & 0.0010\\ 
  $1/f$ noise & 2 & 75.04 & 99.28 & 0.4351 \\
  White noise & 3 & 99.97 & 100.00 & 0.0006 \\ 
  $f^2$ noise & 3 & 99.97 & 100.00 & 0.0006 \\ 
  $1/f$ noise & 3 & 75.95 & 99.32 & 0.4452 \\
  White noise & 5 & 99.95 & 100.00 & 0.0010 \\ 
  $f^2$ noise & 5 & 99.94 & 100.00 & 0.0012 \\ 
  $1/f$ noise & 5 & 76.65 & 99.54 & 0.4448 \\
  \\
  \bottomrule[1.25pt]
  \end{tabular}
}
\end{table}

We found in Table \ref{tab_singlePeak_robustness} that $\HILDE$ is very robust against the additional $f^2$ but effected by $1/f$ noise. At presence of the latter noise, the standard deviation estimation on the long segments is wrong which causes the detection of false positives in roughly a quarter of the cases. Note, that false positives are caused by the underestimated standard deviation but also by the long range dependency itself. However, the false positives have a small amplitude and therefore do not influence the analysis severely or can be removed by postfiltering. Parameter estimation (not displayed) is slightly effected by $1/f$ noise (estimation of the change-point locations is slightly worse, but estimation of the size of the change is even improved), but not affected by presence of $f^2$ noise.

\section{Data analysis}\label{sec:analysis}
\subsection{Measurements}
We analyze single channel recordings of PorB from \textit{Neisseria meningitidis} (recall the last paragraph in the introduction). In the following we analyze six traces, each of them is one minute long and consists of $600\,000$ observations. An example is shown in Figure \ref{fig:PorBHeteroData}, which shows distinct heterogeneous noise.\\
Measurements were performed on solvent-free planar bilayers using the Port-a-Patch (Nanion Technologies). Giant unilamellar vesicles (GUVs) composed of 1,2-diphytanoyl-\textit{sn}-glycero-3-phosphocholine (DPhPC)/cholesterol (9:1) were prepared by electroformation (AC, U = 3 V, peak-to-peak, f = 5 Hz, t = 2 h) in the presence of 1 M sucrose at \SI{20}{\celsius}. Spreading of a GUV in $\SI{1}{\Molar}$ KCl, $\SI{10}{\milli\Molar}$ HEPES, pH 7.5 on an aperture (d = 1-5 $\mu$m) in a borosilicate chip by applying 10-40 mbar negative pressure resulted in a solvent-free membrane with a resistance in the G$\Omega$ range. Once the membrane with a G$\Omega$ seal was formed, varying amounts of a PorB stock solution (2.2 $\mu$M in 200 mM NaCl, 20 mM Tris, 0.1\% (w/w) LDAO, pH 7.5) were added to the buffer solution (50 $\mu$L) at an applied DC potential of +40 mV. Current traces were recorded at a sampling rate of $\SI{10}{\kilo\hertz}$ and filtered with a low-pass four-pole Bessel filter of $\SI{1}{\kilo\hertz}$ using an Axopatch 200B amplifier (Axon Instruments). For digitalization, an A/D converter (Digidata 1322; Axon Instruments) was used.

\subsection{Idealization}
Idealizations are obtained by $\HILDE$ with parameter choices as in Section \ref{sec:parameterchoices}. Moreover, an illustrative comparison with other approaches was discussed in the introduction (recall Figures \ref{fig:PorBHeteroHMM}-\ref{fig:PorBHeteroHSMUCE}). In Figure \ref{fig:PorBHeteroData} we see that the channel switches frequently between two conductance levels, roughly between \SI{0.04}{\nano\siemens} and \SI{0.36}{\nano\siemens}, the variance is roughly $\SI[parse-numbers = false]{6.1 \cdot 10^{-5}}{(\nano\siemens)^2}$ in the closed state and $\SI[parse-numbers = false]{10^{-3}}{(\nano\siemens)^2}$ in the open state. Moreover, several artifacts seem to be present, see for instance the fluctuating conductance in the open state in the first ten seconds. We stress that such artifacts heavily disturb any idealization that assumes a HMM, confer Figure \ref{fig:PorBHeteroHMM}. Contrarily, the model-free idealization by $\HILDE$ (Figure \ref{fig:PorBHeteroJILTAD}) recovers all visible features on small as well as on large temporal scales accurately. In particular, the zooms into single peaks (Figure \ref{fig:PorBHeteroJILTAD}, lower panels) shows that $\HILDE$ fits the observations well which is also a confirmation of our model. Since PorB forms three pores, four different conductance levels are possible. However, in this measurement we see only two different conductance levels. Such a cooperative opening and closing was observed before, see for instance \citep{song1998successful}.

\subsection{Analysis of flickering dynamics}\label{sec:analysisGating}
\begin{figure}[!htb]
\centering
\includegraphics[width = 0.4\textwidth]{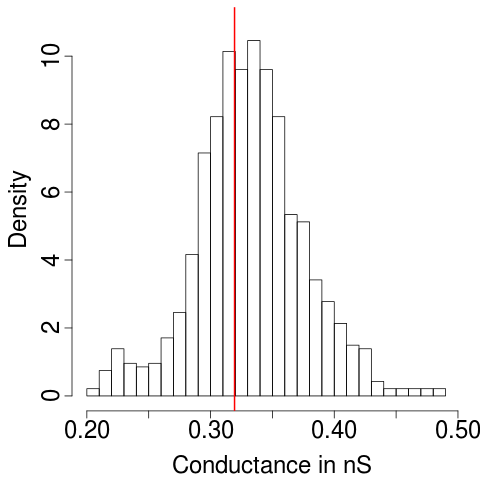}
\caption{Histograms of the amplitudes between $\SI{0.2}{\nano\siemens}$ and $\SI{0.5}{\nano\siemens}$. The vertical red line indicates the estimated amplitude of \SI{0.3194}{\nano\siemens} by the half sample mode.}
\label{fig:PorBheteroAmplitude}
\end{figure}

We now use the obtained idealizations to analyze the gating dynamics in a similar fashion as the simulated data in Section \ref{sec:hmm}. We will focus in this section on $\HILDE$, but we will compare it in Section \ref{sec:appendixanalysisothers} in the supplement with analyses based on $\JULES$, $\HSMUCE$ and $\HMM$. We say a channel \textit{opens} (a gating event from the lower conductance level to the higher conductance level) if the idealized level is between \SI{0.25}{\nano\siemens} and \SI{2}{\nano\siemens} and the previous level is between \SI{0}{\nano\siemens} and \SI{0.1}{\nano\siemens}. To study the amplitude, we consider the conductance difference of all such events. Figure \ref{fig:PorBheteroAmplitude} shows a histogram of the so obtained amplitudes between \SI{0.2}{\nano\siemens} and \SI{0.5}{\nano\siemens}. All other events are either closing events or are considered as artifacts. Such artifacts can for instance be base line fluctuations as discussed in the introduction. We stress that an analysis of the closing events leads to very similar results.

\begin{figure}[!htb]
\centering
\begin{subfigure}{0.32\textwidth}
\includegraphics[width = \textwidth]{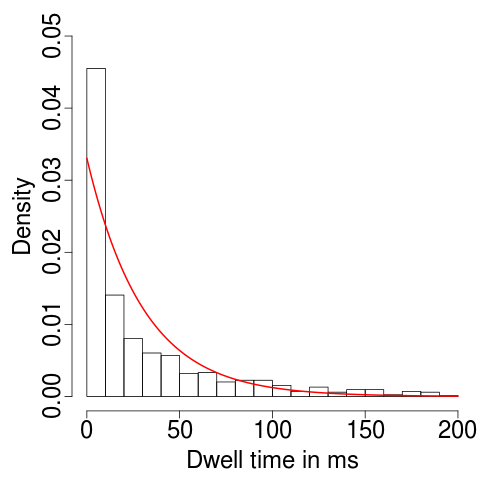}
\subcaption{All events between $\SI{0.1}{\milli\second}$ and $\SI{200}{\milli\second}$.}
\label{subfig:A:dwellPorBhetero_JILTAD}
\end{subfigure}
\begin{subfigure}{0.32\textwidth}
\includegraphics[width = \textwidth]{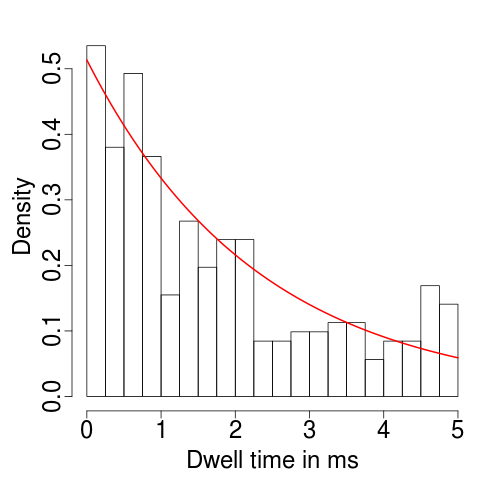}
\subcaption{Short events between $\SI{0.1}{\milli\second}$ and $\SI{5}{\milli\second}$.}
\label{subfig:A:dwellshortPorBhetero_JILTAD}
\end{subfigure}
\begin{subfigure}{0.32\textwidth}
\includegraphics[width = \textwidth]{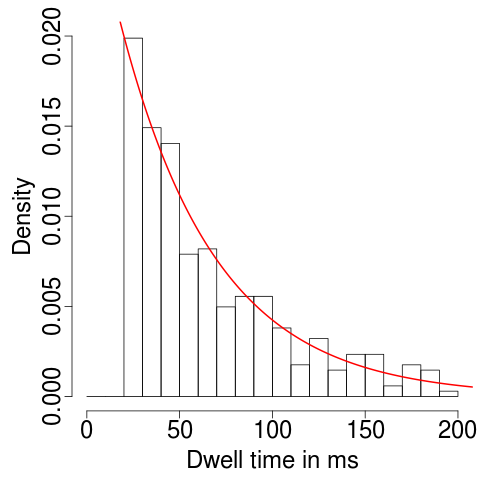}
\subcaption{Long events between $\SI{20}{\milli\second}$ and $\SI{200}{\milli\second}$.}
\label{subfig:A:dwelllongPorBhetero_JILTAD}
\end{subfigure}
\caption{Histograms of the dwell times in the open state of all opening events with amplitude between $\SI{0.2}{\nano\siemens}$ and $\SI{0.5}{\nano\siemens}$ together with exponential fits using a missed event correction (red line).}
\label{fig:dwellPorBhetero}
\end{figure}

The histogram in Figure \ref{fig:PorBheteroAmplitude} shows only one mode. Hence, all events have the same amplitude up to measurements and idealization errors. This means especially that also the flickering events are full-sized. An amplitude of \SI{0.3194}{\nano\siemens} is estimated by the half sample mode \citep{robertson1974iterative}, computed in \texttt{R} by using the \textit{modeest} package. Note that other mode estimators or Gaussian mean estimation lead to similar results. This amplitude coincides with the one obtained by a manual analysis using the pClamp 10.2 software package (Axon Instruments), see \citep{Bartschetal18}.\\
We now analyze the dwell time in the open state and how frequently the channel opens. We take into account events with an amplitude between $\SI{0.2}{\nano\siemens}$ and $\SI{0.5}{\nano\siemens}$ and with a dwell time between $\SI{0.1}{\milli\second}$ and $\SI{200}{\milli\second}$, since shorter events cannot be detected reliably and longer events are rare and often interrupted by artifacts. Histograms of the dwell time in the open state are shown in Figure \ref{fig:dwellPorBhetero} together with an exponential fit using a missed event correction like in \citep{Peinetal18}.

\begin{figure}[!htb]
\centering
\includegraphics[width = 0.5\textwidth]{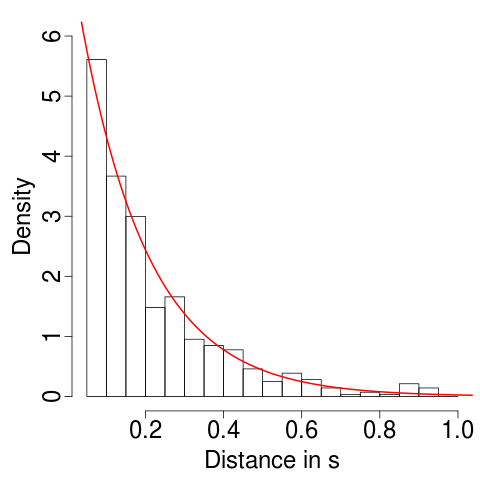}
\caption{Histograms of the distances between two opening events with amplitude between $\SI{0.2}{\nano\siemens}$ and $\SI{0.5}{\nano\siemens}$ together with exponential fits using a missed event correction (red line).}
\label{fig:distancePorBhetero}
\end{figure}

Interestingly, the dwell times do not fit a single exponential distribution, but when we split the events in short (shorter than \SI{5}{\milli\second}) and long (longer than \SI{20}{\milli\second}) ones, both fit exponential distributions very well, with an estimated average duration of \SI{51.62}{\milli\second} and \SI{2.31}{\milli\second}, respectively. Note, that these estimations are approximations, since an exponential distribution with a large / small rate generates with a small probability a long / short event, but since the average dwell times are very different this error is negligible. To best of our knowledge, fast and slow gating at the same time was not observed for PorB before. However, \citet{Grosse.etal.14} showed that the loop within the pore structure of OmpG leads to fast flickering (fast time constant). If the loop is removed, there is still gating observed but less frequent (slower time constant). Even though this is not the same protein, in PorB we have a loop L3 which is also localized in the pore and forms an $\alpha$-helix in its center, which constricts the pore to its narrowest point. Hence, our findings support that similar dynamics might occur for PorB as well.\\
We are now analyzing the distance between two opening events. This is shown in Figure \ref{fig:distancePorBhetero}. Moreover, we analyze how many of the openings are short or long. Once again we apply a correction for missed events.

The distance between two events seem to be exponentially distributed and the estimated rate is \SI{5.75}{\hertz}. We found that $39.08\%$ of all opening events were short events. Moreover, we found in Section \ref{sec:appendixanalysisothers} in the supplement that all results obtained by $\HILDE$ could be confirmed by at least one other approach, but none of the other methods was able to reproduce all results obtained by $\HILDE$.

\section{Discussion and Outlook}\label{sec:discussion}
In this paper we proposed a new model-free idealization method for ion channel recordings, called $\HILDE$. In comparison to existing approaches, $\HILDE$ provides still reasonable idealizations under heterogeneous noise, for instance caused by open channel noise. Moreover, it detects and idealizes flickering events reliable, is fully-automatic and can be computed efficiently. It offers great flexibility in adapting to the needs of a specific data analysis by modifying the error probabilities $\alpha = \alpha_1 + \alpha_2$ and the scale $l_{\max}$ that distinguishes short and long events. Its precise idealization is confirmed by simulations and a real data application to PorB recordings. We found that these recordings contain opening events of significantly different length.\\
We stress that $\HILDE$ is modular, i.e., single components like the choice of the test statistics and functionals to optimize can be changed without further modifications. This can be used to adapt $\HILDE$ to specific challenges in the measurements. We will discuss several such possibilities in the following. Some of them are implemented in the \textit{clampSeg} package and just require to choose different parameters, for others few lines of code have to be modified.

\subsection{Alternative approaches}\label{sec:discussionalternatives}
A different underlying interval set can be used for the multiresolution test. The set of all intervals of dyadic length provides in general a good compromise between detection power and computation time. But, if a larger detection power is required, the set of all intervals can be used at the price of a larger computational complexity. The other way around, if faster computation is demanded, a smaller interval set, for instance the dyadic partition like in \citep{Pein.Sieling.Munk.16}, can be used. This might be particularly beneficial in situations in which the multiresolution test detects almost no events which results in a large computation time. Interesting alternatives are also the approaches in \citep{chan2013detection, kovacs2020seeded} which require only a slightly larger computational effort than the use of all intervals of dyadic length but detects change-points in a certain sense statistically optimally. A different way to increase the detection power is to use likelihood ratio tests, again at computational expenses. We found in simulations (not displayed) that the likelihood ratio test statistic is slightly more powerful on small scales, but much slower to compute. However, a slightly worse detection power on small scales should not be a big concern, since a refinement by local tests will be done in the next step. Also for detecting events on small scales by local tests, see Section \ref{sec:smallscales}, different statistics can be used to increase the detection power. For instance the likelihood ratio test or maximum likelihood estimators for the parameters $(\valmu,s^2)$ can be considered. However, they are computationally very demanding, since the likelihood function involves the inverse and the determinant of the covariance matrix given by \eqref{eq:covpeak}.\\%
Finally, our deconvolution approach assumes still homogeneous noise which we found in simulations works still well at presence of open channel noise, see Tables \ref{tab_singlePeak_hetero_estimationLocation} and \ref{tab_singlePeak_hetero_estimationLevel}. Taking into account the heterogeneous noise might be beneficial, in particular if the noise level differences are large, but difficult, maybe even impossible, since avoiding an ill-conditioned matrix by regularization and keeping the variance levels might be impossible to achieve at the same time.

\subsection{Homogeneous noise}\label{sec:discussionhomogeneous}
We designed $\HILDE$ particularly to deal with heterogeneous noise. However, taking into account the convolution explicitly when detecting changes is also beneficial if the noise is homogeneous, i.e., a constant variance is assumed. In this situation, its detection power can be further improved by small modifications that utilize the assumption of a constant variance, they are explained in Section \ref{sec:appendixhomogeneousnoise} in the supplement. We found that $\HILDE$ has a better detection power than $\JULES$ \citep{Peinetal18}, but at the price of worse separation properties and a larger computation time. More precisely, in the simulations in Section IV-B in \citep{Peinetal18} we found that $\JULES$ is able to detect an isolated peak of length $3.1/\sr$ with probability almost one. In comparison, $\HILDE$ requires only $2.3/\sr$ if homogeneous noise is assumed and $2.8/\sr$ if heterogeneous noise is assumed (see Section 3.9.3. in \citep{Pein17}). Remarkably, the detection power of $\HILDE$ is even larger than the one of $\JULES$ if $\HILDE$ does not use the assumption of homogeneous noise which illustrates how much detection power is lost by not taking into account the convolution.

\subsection{Idealizing the variance}
Our focus was on idealizing the conductance while the unknown variance was considered as a nuisance parameter. However, as a byproduct $\HILDE$ can easily be extended to an idealization of the variance which offers for instance a residual analysis of the noise to validate a given model. To this end, we use $\HILDE$ to estimate the change-point location of the conductance and assume that these are the change-points of the variance as well. Note that the model of Section \ref{sec:model} allows the variance to stay constant at such a location but precludes further variance changes. With the definitions from before (see Section \ref{sec:methodology}), if a segment is long, the square of the estimator in \eqref{eq:eststd} can be used. Afterwards, the variance on short segments can be estimated by the estimator in \eqref{eq:estvar}. The resulting function will be an idealization of the variance. 

\bibliographystyle{apalike}
\bibliography{Literature}

\newpage
\pagenumbering{arabic}
\section*{Supplement to\\ Heterogeneous Idealization of Ion Channel Recordings - Open Channel Noise}

\hspace*{10pt}

\section{Large scales}\label{sec:appendixlongsegments}
Following the ideas of $\operatorname{HSMUCE}$ \citep{Pein.Sieling.Munk.16} we propose the idealization $\hat{f}$ by 
\begin{equation}\label{eq:multiscaleestimator}
  \hat{f} := \sum_{k=0}^{\hat{K}}\,\hat{\valmu}_k\,\ind_{[\hat{\tau}_k, \hat{\tau}_{k+1})} 
  := \argmin_{\tilde{f}\in \mathcal{F}_{\hat{K}}, M(\tilde{f})\leq 0}\, \sum_{i=1}^{n}{\frac{\big(Y_i - \tilde{f}(i/n)\big)^2}{\hat{\sigma}_i^2}},
\end{equation}
with $\hat{\sigma}_i=\sum_{j\in I^i}{(Y_j-\tilde{f}(i/n))^2}/\vert I^i\vert$, $I^i=\{j\in\{1,\ldots,n\}\, :\, \tilde{f}(k/n)=\tilde{f}(i/n)\ \forall\ k=j,\ldots,i\}$. In other words, $\hat{f}$ is the maximum likelihood estimator restricted to all solutions of the optimization problem
\begin{equation}\label{eq:optimizationproblem}
\tilde{f}\in \mathcal{F}_{\hat{K}} \text{ such that } M(\tilde{f})\leq 0.
\end{equation}
Thereby, $\mathcal{F}_k$ is the set of all candidate signals in \eqref{eq:signal} with $K=k$ changes, $k\geq 0$. $\hat{K}$ is the estimated number of changes defined by the minimal number $k$ for which there exits an $\tilde{f}\in \mathcal{F}_k$ such that $M(\tilde{f}) \leq 0$. Finally, $M(\tilde{f})$ denotes the multiresolution test statistic
\begin{equation}\label{eq:multistats}
  M(\tilde{f}) := 
 \max_{\substack{1\leq i \leq j \leq n\\
 \tilde{f}([i/\sr,j/\sr]) \text{ const.}}}
 \,\left\{ T_{i,j}(\tilde{f})  - q_{\vert j - i +1\vert}\right\}.
\end{equation}
This \emph{tests simultaneously} on all scales (resolution levels) whether $\tilde{f}$ fits the data well. If it does not, the \textit{local test statistic} $T_{i,j}$ will be larger than the \textit{scale dependent critical value} $q_{\vert j - i +1\vert}$, exact definitions are given below, and $\tilde{f}$ will not be considered as a potential idealization. Such a method has many favorable properties, for more details see \citep{Frick.Munk.Sieling.14, Pein.Sieling.Munk.16, Peinetal18}. Most importantly, the number of false positives is controlled if the scale dependent critical values are defined appropriately (one possible choice is outlined below), see Theorem \ref{theorem:overestimation}.\\
We use as test statistic the statistic in (1.5) in \citep{Pein.Sieling.Munk.16} without taking into account the first $m$ observations (\textit{$\JSMURF$ principle}), i.e.,
\begin{equation}\label{eq:localtesthjsmurf}
T_{i,j}(\tilde{f}):=(j-i+1-m)\frac{(\overline{Y}_{i+m,j} - \tilde{f}_{ij})^2}{2\hat{s}_{i+m,j}^2},
\end{equation}
with $\tilde{f}_{ij}$ the conductance level of $\tilde{f}$ on the interval $[i/n,j/n]$, $\overline{Y}_{i+m,j}=(j-i+1-m)^{-1}\sum_{l=i+m}^{j}{Y_l}$ and $\hat{s}_{i+m,j}=(j-i-m)^{-1}\sum_{l=i+m}^j{(Y_l-\overline{Y}_{i+m,j})^2}$. This statistic estimates the variance locally and is large if the mean of the observations differ significantly from the conductance level $\tilde{f}_{ij}$. The scale dependent critical values $q_{m+1},\ldots,q_n$ are obtained in a universal manner by Monte Carlo simulations as described in Section 2 of \citep{Pein.Sieling.Munk.16} such that \eqref{eq:multistats} is a level $\alpha_1$-test and different scales are balanced by weights. Here, we use the default choice of uniform weights. The error level $\alpha_1\in(0,1)$ has to be fixed in advance (by the experimenter) to control the number of false positives of the estimate $\hat{f}$ as stated in the following Theorem \ref{theorem:overestimation}. See Section \ref{sec:parameterchoices} for a discussion how to choose $\alpha_1$.
\begin{Theorem}\label{theorem:overestimation}
Assume the heterogeneous ion channel model from Section \ref{sec:model} and let $\hat{f}$ be defined as in \eqref{eq:multiscaleestimator} with error level $\alpha_1$. Then, for any $f$ and $\sigma$ in \eqref{eq:signal} the probability that $\hat{f}$ overestimates the number of changes is bounded by $\alpha_1$, i.e.
\begin{equation}
\mathcal{P}\left(\hat{K}>K\right)\leq \alpha_1.
\end{equation}
\end{Theorem}
To keep the method computationally feasible, we only evaluate the maximum in \eqref{eq:multistats} over the system of all intervals that contain a dyadic number of observations, i.e., the maximum in \eqref{eq:multistats} is only taken over all $1\leq i \leq j\leq n$ such that $j - i + 1 = 2^l$ for some $l\in\N_0$. This reduces the complexity of the system from $\mathcal{O}(n^2)$ to $\mathcal{O}(n\log(n))$ intervals. This speeds up the pruned dynamic program and the simulations in Section \ref{sec:simulations} show that this interval set is large enough to allow a good performance. More details and a discussion of the run time are given in Section \ref{sec:computation}. Note that \citep{Hotz.etal.13} performed tests on a slightly different interval set. They required that $j - i + 1 - m$ is a dyadic number. Although depending on the true signal, their choice leads in general to a better detection power on scales slightly larger than the filter length $m/\sr$, but its computation lasts much longer. And missed short events can be detected by the upcoming local tests. The very long computation time was one major criticism in \citep{gnanasambandam2017unsupervised}. To be fair, both implementations differ in other points, too, and a fast implementation of their interval set might be possible as well, but our approach was easier to integrate in the dynamic programming framework of the \textit{stepR} package \citep{stepR}. Finally, we remark that the restricted maximum likelihood estimator ignores the convolution and hence the locations of the detected changes are typically a little bit shifted to the right which will be corrected in the upcoming deconvolution step in Section \ref{sec:deconvolution}. Different to this, \citep{Hotz.etal.13} suggested to move all locations by a constant factor $t_0$, only depending on the filter, to the left, but our deconvolution step will (usually) be more precise.\\

\section{Small scales}\label{sec:appendixshortsegments}
The upcoming three paragraphs describe precisely how the local tests are performed to detect short events that are missed in the idealization obtained in Section \ref{sec:large scales}.

\paragraph*{Obtaining the hypotheses and alternatives}
In this paragraph we give details for the construction of the hypothesis \eqref{eq:hypothesis} and alternative \eqref{eq:alternative}. Assuming the model from Section \ref{sec:model}, we see that the expectation of the observations $Y_i,\ldots,Y_j$ is determined by the signal on the interval $[(i-m)/\sr,j/\sr]$. The other way around, information about the underlying signal on an interval $[i/\sr,j/\sr]$ is provided by the observations $Y_{i+1},\ldots,Y_{j+m-1}$ and the signal on $[(i-m+1)/\sr, (j+m-1)/\sr]$ effects the expectation of these observations. Hence, for a local test on an interval $[i/\sr,j/\sr]$ we distinguish few scenarios depending on how many changes the previous idealization has in $[(i-m+1)/\sr, (j+m-1)/\sr]$.\\
If no change is contained, we test a constant signal against the alternative of an additional event on $[i/\sr,j/\sr]$ with an arbitrary conductance level. If one change is contained, we test a signal with one change, exact details are discussed below, against the alternative of an additional event on $[i/\sr,j/\sr]$ with an arbitrary conductance level. If the test rejects, the single change is replaced by two and the exact locations and the conductance level between these two changes are obtained in the upcoming deconvolution step in Section \ref{sec:deconvolution}. In the rare situation that two or more changes are present no local test is performed to save computation time, since the parameters of more than two changes can anyway not be estimated in the upcoming deconvolution step. Moreover, we only test on intervals with start and end point at the observation grid. Both limitations can be narrowed as discussed in Section \ref{sec:discussionalternatives}, at the price of a larger computation time. But, we found that our choices are sufficient for the data we analyze.\\
We now describe how we obtain the parameters $\tau, \valmu_L, \valmu_R, s_L$ and $s_R$ in the hypotheses in \eqref{eq:hypothesis} and alternatives in \eqref{eq:alternative}. To this end, note that changes in the idealization from Section \ref{sec:large scales} are typically slightly shifted to the right, since the convolution is ignored and the lowpass filter acts only in the past. Hence, if we simply obtain the parameters from the previous idealization, many hypotheses will be wrongly rejected, even if the true underlying signal has only one change in $[(i-m+1)/\sr, (j+m-1)/\sr]$. To correct for this, we reestimate the locations of all isolated changes by deconvolution. This is performed locally as described in the upcoming Section \ref{sec:deconvolution}. Since we assume for testing that all changes are on the observation grid, we perform the deconvolution only at the observation grid, without any refinement at finer scales. This includes a reestimation of the conductance levels on long segments by medians. In other words, as the hypothesis we assume the signal which will be obtained by deconvolution if no test rejects, up to refinements using finer grids. At the same time, estimation of the conductance levels on long segments by the median guarantees that they are not too badly estimated even if few short peaks are missed.\\
On long segments, in addition to the expectation, the standard deviation $s$ is estimated by 
\begin{equation}\label{eq:eststd}
\hat{s}(Y_a,\ldots,Y_b)=\frac{\operatorname{IQR}\left(Y_{a+m}-Y_{a},\ldots, Y_b-Y_{b-m}\right)}{2\Phi^{-1}(0.75)\sqrt{2(\kernel \ast\kernel)(0)}},
\end{equation}
using the same observations as used for estimating the expectation. Here, $\Phi^{-1}$ denotes the quantile of the standard normal distribution.\\
Finally, we recommend to choose $l_{\max}$ such that events on all larger scales are already detected by the previous idealization (or have such a small jump size that they are also not detectable by the tests in this step). We found in simulations (not displayed) that $l_{\max}=65$ is a suitable default choice.

\paragraph*{Local testing}
In this paragraph we propose a test that provides a good trade-off between detection power for events in the measurements in Section \ref{sec:analysis} and computational complexity. We start with estimating the unknown parameters $\valmu$ and $s$ under the alternative.\\
To estimate $\valmu$  we use the least squares estimator
\begin{equation}\label{eq:estmeanls}
\begin{split}
\hat{\valmu}&:=\argmin_{\valmu\in\R}{\sum_{l=i+1}^{j+m-1}{\big(Y_l-\E[Y_l]\big)^2}}=\argmin_{\valmu\in\R}{\sum_{l=i+1}^{j+m-1}{\big(Y_l-v_l\valmu-\valmu_{LR,l}\big)^2}}\\
&=\frac{\sum_{l=i+1}^{j+m-1}{v_l(Y_l-\valmu_{LR,l})}}{\sum_{l=i+1}^{j+m-1}{v_l^2}},
\end{split}
\end{equation}
where
\begin{equation}\label{eq:v}
v_l:= \mathcal{F}_m(l/\sr - \tau_L) - \mathcal{F}_m(l/\sr - \tau_R)
\end{equation}
and
\begin{equation}\label{eq:valmuLR}
\valmu_{LR,l}:=\valmu_L [1-\mathcal{F}_m(l/\sr - \tau_L)] + \valmu_R \mathcal{F}_m(l/\sr - \tau_R),
\end{equation}
with $\mathcal{F}_m$ the antiderivative (step function) of the truncated filter kernel. Moreover, it follows from \eqref{eq:cov} that under the alternative \eqref{eq:alternative} the covariance is given by
\begin{equation}\label{eq:covpeak}
\COV\big[Y_l,Y_{l+r}\big]=\left\{\begin{array}{cl}
w_{l,r}s^2 + s_{LR,l,r}^2\, & \text{for } \vert r \vert=0,\ldots,m,\\
0 & \text{for } \vert r\vert > m,
\end{array}\right.
\end{equation}
with
\begin{equation}\label{eq:wij}
w_{l,r} := \mathcal{A}_m(l/\sr - \tau_{L}, r/\sr) - \mathcal{A}_m(l/\sr - \tau_{R}, r/\sr)
\end{equation}
and
\begin{equation}\label{eq:varij}
s_{LR, l,r}^2  := s_L^2\big[\mathcal{A}_m(\infty, r/\sr) - \mathcal{A}_m(l/\sr - \tau_{L}, r/\sr)\big] + s_R^2 \mathcal{A}_m(l/\sr - \tau_{R}, r/\sr).
\end{equation}
Hence, for estimating the variance $s^2$ we use the weighted estimator
\begin{equation}\label{eq:estvar}
\hat{s}^2:=\max\left(\frac{\sum_{l=i+1}^{j+m-1}{w_{l,0}\big(Y_l-v_l\hat{\valmu}-\valmu_{LR,l}\big)^2}-B(s_L, s_R)}{A},0\right),
\end{equation}
with $A$ and $B(s_L, s_R)$ such that
\begin{equation}\label{eq:estvarepectation}
\E\left[\sum_{l=i+1}^{j+m-1}{w_{l,0}\big(Y_l-v_l\hat{\valmu}-\valmu_{LR,l}\big)^2}\right]=:A s^2+B(s_L, s_R).
\end{equation}
Note that the random variable of which we take the expectation in \eqref{eq:estvarepectation} can be written as a quadratic form $(Y_{i+1,j+m-1}-\E[Y_{i+1,j+m-1}])^t C (Y_{i+1,j+m-1}-\E[Y_{i+1,j+m-1}])$, where $Y_{i,j}=(Y_i,\ldots,Y_j)^t$ and all entries of the matrix $C$ are non-negative and depend only on $v_l$ and $w_{l,0}$, $l=i+1,\ldots,j+m-1$. This combined with \eqref{eq:covpeak} confirms the proposed structure in \eqref{eq:estvarepectation} follows and allows to computed $A$ and $B(s_L, s_R)$ explicitly.\\
Note that the estimator $\hat{\valmu}$ is unbiased, while for $\hat{s}$ this would be true without the projection of negative values to zero in \eqref{eq:estvar}, which however reduces the mean square error.\\
Using these estimators, under the alternative the observation $Y_l$ has estimated expectation $\hat{\valmu}_{1,l}:=v_l\hat{\valmu}+\valmu_{LR,l}$ and estimated variance $\hat{s}_{1,l}^2:=w_{l,0}\hat{s}^2 + s_{LR,l,0}^2$. Under the null hypothesis the observation $Y_l$ has expectation $\valmu_{0,l}:=\valmu_L\big[1-\mathcal{F}_m(t-\tau)\big]+\valmu_R \mathcal{F}_m(t-\tau)$ and variance $s_{0,l}^2:= s_L^2 \big[\mathcal{A}_m(\infty,0)-\mathcal{A}_m(l/\sr-\tau)\big]+s_R^2 \mathcal{A}_m(l/\sr-\tau, 0)$. Finally, using these estimators we propose the test statistic
\begin{equation}\label{eq:testimprove2param}
T_i^j:=\sum_{l=i+1}^{j+m-1}{\log\left(\frac{s_{0,l}^2}{\hat{s}_{1,l}^2}\right) + \frac{(Y_l-\valmu_{0,l})^2}{s_{0,l}^2} - \frac{(Y_l-\hat{\valmu}_{1,l})^2}{\hat{s}_{1,l}^2}}.
\end{equation}
We are aware that this test statistic and its underlying estimators might be improvable with respect to efficiency of the estimators and the power of the resulting test for its various alternatives, for a more detailed discussion and potential alternatives see Section \ref{sec:discussion}. But, as stressed before, we aimed for a test that has at least a good power for the recordings in Section \ref{sec:analysis} and can be computed efficiently. This will be confirmed by the simulations in Section \ref{sec:simulations}.\\
Moreover, note that this test uses information provided by potential standard deviation changes as well. This is different to the multiresolution test we used in Section \ref{sec:large scales} to detect events on large scales and to $\operatorname{HSMUCE}$. Note that, different to similar ideas in these settings, it can be computed efficiently, since only testing is required and not regression based on these tests. This is another gain of the three step procedure we propose in this paper. The test problem is also of a different type than the one in \citep{Enikeevaetal15}, since we allow the standard deviation to be constant ($s=s_L$ or $s=s_R$) when the conductance changes.

\paragraph*{Critical values}
It remains to choose critical values that balance the different tests appropriately. To this end, we obtain again scale depend critical values by using the approach from Section 2 in \citep{Pein.Sieling.Munk.16}. We apply it with significance level $\alpha_2$ and equal weights $\beta_1,\ldots,\beta_{l_{\max}}=1/l_{\max}$. By this we aim to control the overall probability of detecting an false positive by $\alpha:=\alpha_1 + \alpha_2$. While we showed in Theorem \ref{theorem:overestimation} such a control for the multiresolution procedure in Section \ref{sec:large scales}, we are not able to prove such a bound for the local tests as well, since the previous idealization might not be exactly the true signal up to events on shorter temporal scales and hence the observations are not generated exactly according to the hypotheses \eqref{eq:hypothesis} and alternatives \eqref{eq:alternative}. Moreover, to speed up the required Monte-Carlo simulations we use the following simplification. When computing the test statics in the Monte-Carlo simulations, we ignore the previous idealization step and assume instead a constant signal. Since the idealization by $\JSMURF$ leads with probability at least $1-\alpha_1$ to a constant idealization, this error is negligible. All in all, we found in simulations that the local tests keep the error level $\alpha_2$ well.

\paragraph*{Multiple dependent rejections}
Usually one event in the data causes rejections of multiple tests. Hence, we only add the event that corresponds to the rejection with the largest test statistic among all rejections on intervals that intersect or adjoin each other. More precisely, two rejections on intervals $[i_1/\sr,j_1/\sr]$ and $[i_2/\sr,j_2/\sr]$ are only considered as two separated events if the intervals are disjoint and (w.l.o.g. let $j_1 < i_2$) there exists an $l\in\{j_1+1,\ldots,i_2-1\}$ such that all tests on intervals containing $l/\sr$ accept the hypothesis. The choice to consider the rejection with the largest test statistic is a natural choice for all tests on intervals of the same length, since they share the same distribution (under their respective null hypotheses and alternatives). For tests on intervals of different lengths this is not exactly true. Nonetheless, we found that considering the rejection with the largest test statistic works very well in practice. That is because usually the test statistics are much larger when their alternative is true than when their null hypothesis is true, which outweighs the (slightly) different distributions (under their respective null hypotheses and alternatives). Also note that a slight missestimation of a location does not have a noticeable effect, since the final estimation of them is obtained in the upcoming deconvolution step.

\section{Local deconvolution}\label{sec:appendixdeconvolution}
In this section we describe two minor modifications on the local deconvolution approach of \citep{Peinetal18} which we made to adapt to the heterogeneous noise setting. As summarized at the end of the introduction the local deconvolution approach of \citep{Peinetal18} requires that two short events are separated by at least one long event. Parameters on long events can be estimated without deconvolution. In our setting we have to estimate mean and standard deviation instead of only the mean as in \citep{Peinetal18}. Hence, our first modification is that we require at least $25$ instead of ten observations in the definition of a long segment to guarantee a reasonable well parameter estimation. Secondly, we adapt the choice of the grids. For the idealization in Section \ref{sec:large scales} and for the detection step of $\JULES$ \citep{Peinetal18} the locations of the estimated changes are shifted to the right, since the convolution was ignored. Hence, we use for a change $\hat{\tau}$ detected in Section \ref{sec:large scales} (and not replaced by two detected changes by the local tests) still the grid $\{\hat{\tau}-m/\sr,\hat{\tau}-(m+1)/\sr,\ldots,\hat{\tau}\}$. However, for a change $\hat{\tau}$ detected by the local tests in Section \ref{sec:smallscales} we use instead $\{\hat{\tau} - \lceil m/2\rceil /\sr,\ldots,\hat{\tau} + \lceil m/2\rceil /\sr\}$, since the locations are not estimated precisely, but also not systematically shifted to one side. The reestimation of the conductance levels on long segments is adapted in the same way. Everything else is performed in the same way as explained in Section 3.2 of \citep{Peinetal18}.

\section{Idealizations by existing approaches}\label{sec:appendixexamples}
In this section we discuss in more detail than in the introduction the idealization of the observations in Figure \ref{fig:PorBHeteroData} by various approaches. We begin with approach that assume a hidden Markov model. We observed two different conductance levels and hence started with two states. We used the following starting values
\[\mu = (0.05, 0.38)^T,\ s = (0.0112, 0.0497)^T,\ P = \left( \begin{matrix} 0.95 & 0.05\\ 0.05 & 0.95 \end{matrix} \right)\]  
for mean, standard deviation and and transition matrix, respectively. The standard deviations were determined by taking the standard deviation of all observations below and above $0.2$, respectively.

\begin{figure}[!htb]
\centering
\includegraphics[width = 0.99\textwidth]{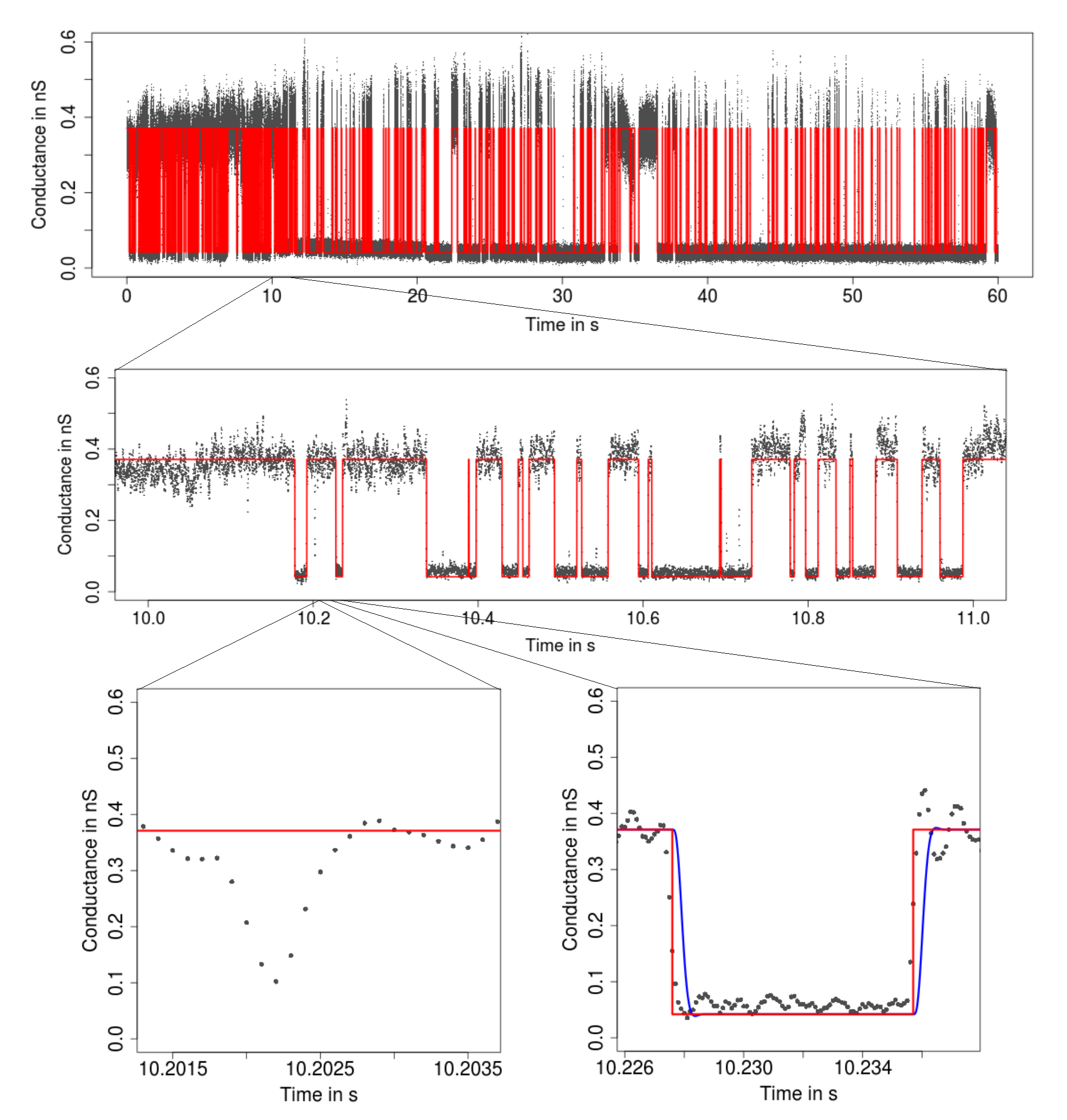}
\caption{Idealization (red) of the data in Figure \ref{fig:PorBHeteroData} assuming a HMM with two states displayed on three different temporal scales. In the lower panels we also show the convolution of the idealization with the lowpass filter (blue). It fits most part of the data well, but misses short events, for instance the event displayed in lower left panel.}
\label{fig:PorBHeteroTwoStates}
\end{figure}

Its idealization, obtained by the Viterbi algorithm and displayed in Figure \ref{fig:PorBHeteroTwoStates}, looks well on all larger temporal scales, but misses short events, for instance the event displayed in the lower left panel. To fit such events well, we added a third state and used the starting values
\begin{equation}\label{eq:startingValuesThreeStates}
\mu = (0.05, 0.05, 0.38)^T,\ s = (0.0112, 0.0112, 0.0497)^T,\ P = \left( \begin{matrix} 0.99999 & 0 & 0.00001\\ 
0 & 0.999 & 0.001\\ 0.02 & 0.03 & 0.95 \end{matrix} \right).
\end{equation}

\begin{figure}[!htb]
\centering
\includegraphics[width = 0.99\textwidth]{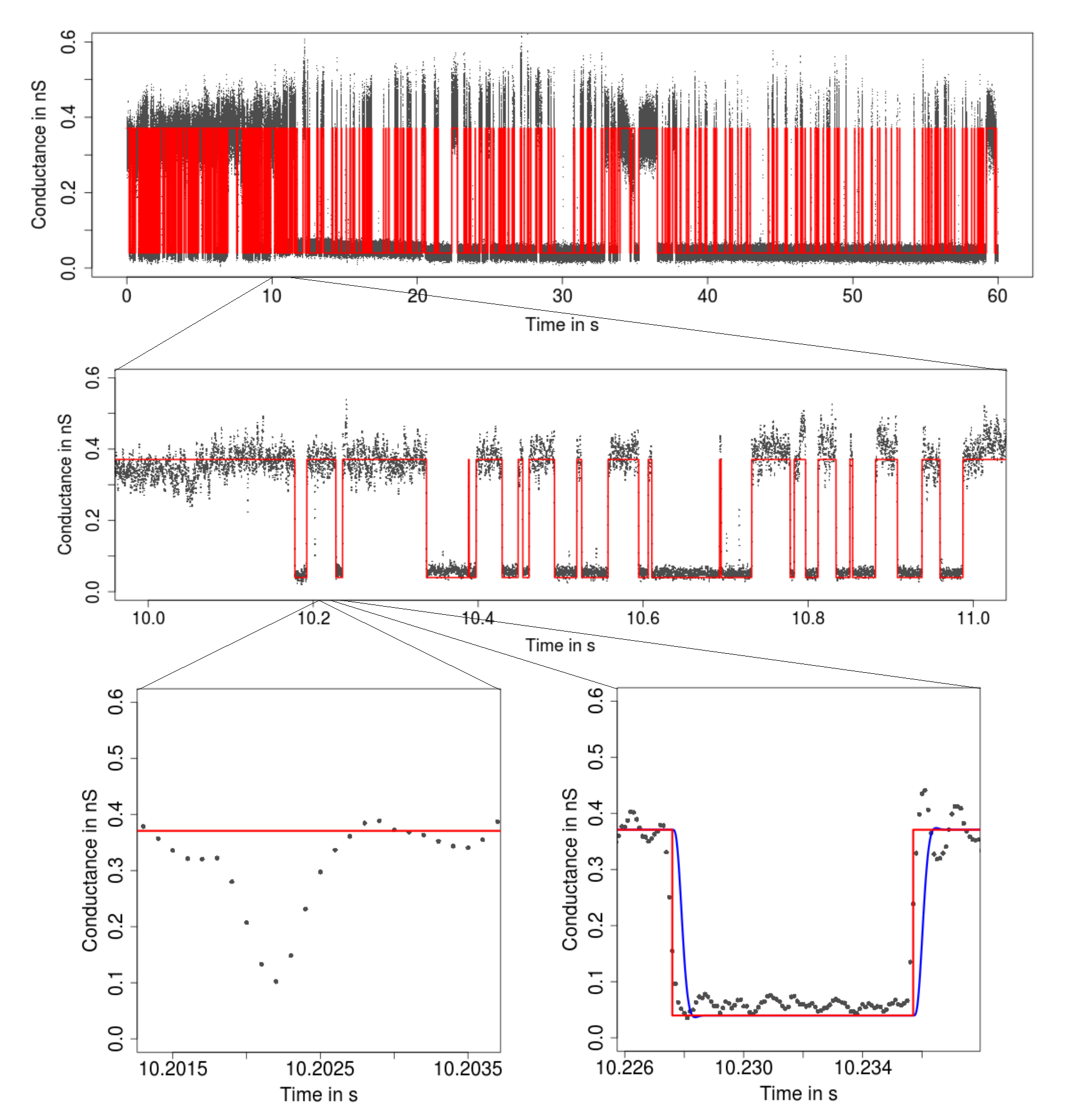}
\caption{Idealization (red) of the data in Figure \ref{fig:PorBHeteroData} assuming a HMM with three states displayed on three different temporal scales. In the lower panels we also show the convolution of the idealization with the lowpass filter (blue). It fits most part of the data well, but misses short events, for instance the event displayed in lower left panel.}
\label{fig:PorBHeteroThreeStates}
\end{figure}

\begin{figure}[!htb]
\centering
\includegraphics[width = 0.99\textwidth]{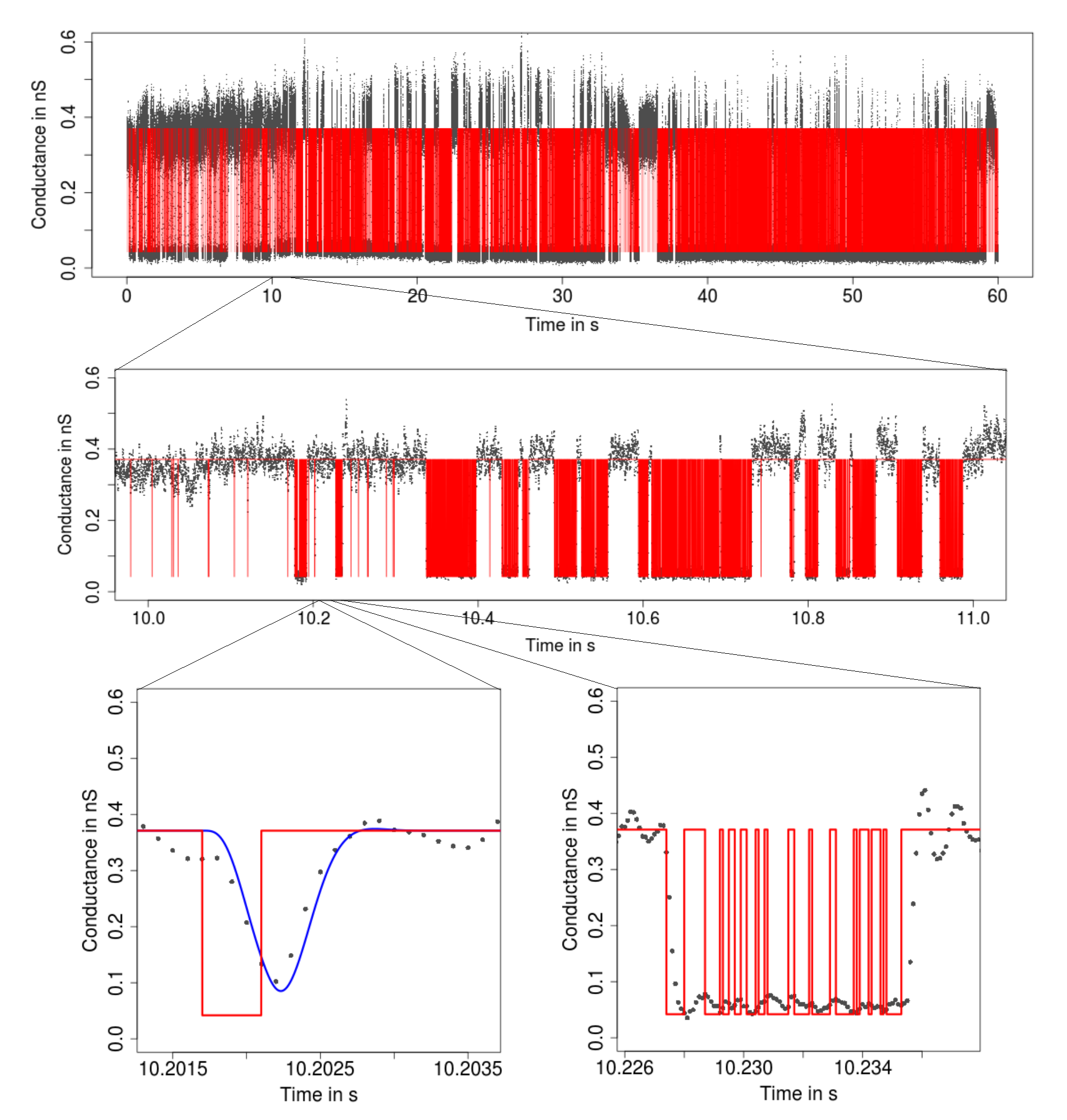}
\caption{Idealization (red) of the data in Figure \ref{fig:PorBHeteroData} assuming a HMM with filtering \citep{Diehn17} displayed on three different temporal scales. In the lower panels we also show the convolution of the idealization with the lowpass filter (blue). It detects very short events but finds a huge number of (most likely false positive) events at parts of low conductance (see for instance the lower right panel).}
\label{fig:PorBHeteroHMM}
\end{figure}

However, the resulting idealization, displayed in Figure \ref{fig:PorBHeteroThreeStates}, did not change much. In fact, the new state is attained only two times, i.e., fits artifacts instead of the short events. We varied the starting values but without much success. To improve results we decided to assume the same expectation and variance for those two states and we used again the starting values in \eqref{eq:startingValuesThreeStates}. Note that this model-class was motivated by the results we obtained from using our model-free idealization approach $\HILDE$ which illustrates nicely how model-free approaches can be used to support HMMs. The resulting idealization, displayed in Figure \ref{fig:PorBHeteroCoupled} in the introduction, however still misses short events. To detect such events we think that it requires to take into account the filtering explicitly. This can be done by introducing so called meta-states \citep{venkataramanan1998identification, venkataramanan1998, deGunst.etal.01}. We used the implementation in \citep{Diehn17}, but found it way to slow to run it for three states. Hence, we only assumed two states but with filtering. We determined the conductance and variance levels as well as a starting value for the transition matrix by fitting an unfiltered HMM. Secondly, we applied an Baum-Welch algorithm, which takes into account a discretized filter, to estimate the transition matrix. It estimates a transition probability from the closed to the open state of $1.66\%$ and of $4.64\%$ for a transition from the open to the closed state. Finally, the observations are idealized by a Viterbi algorithm assuming discrete filtering as well. This idealization is displayed in Figure \ref{fig:PorBHeteroHMM}.

\begin{figure}[!htb]
\centering
\includegraphics[width = 0.99\textwidth]{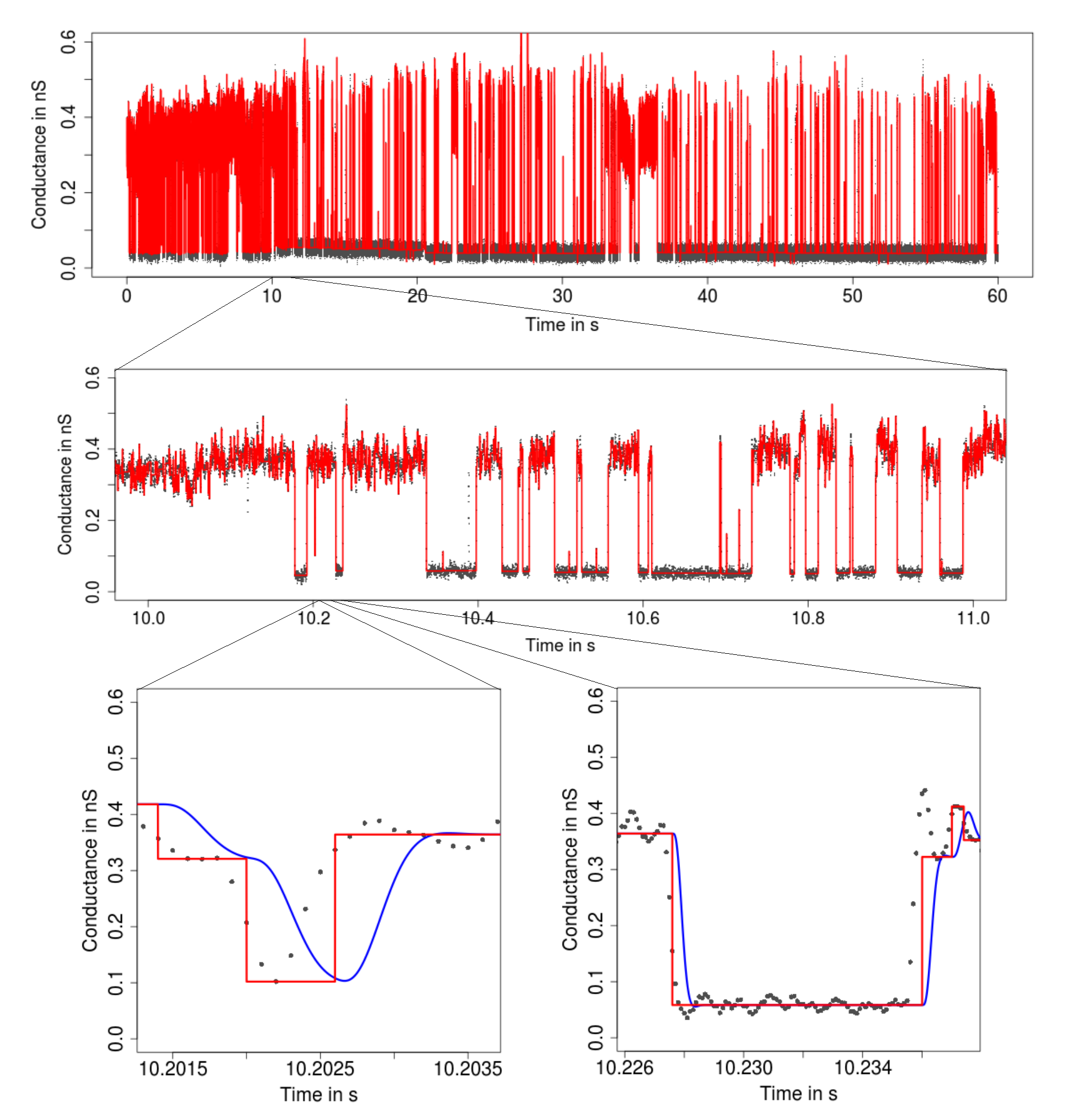}
\caption{Idealization (red) of the data in Figure \ref{fig:PorBHeteroData} by $\TRANSIT$ displayed on three different temporal scales. In the lower panels we also show the convolution of the idealization with the lowpass filter (blue). It detects short events, but finds many small events (which are most likely false positives) at parts of high conductance and high variance (see for instance the idealization of the observations around \SI{0.36}{\nano\siemens} in the middle panel). These detections hinder the decovolution (see for instance the lower left panel) and make the idealization unreliable.}

\caption{Idealization by $\TRANSIT$ (red) of the signal underlying the observations in Figure \ref{fig:PorBHeteroData} and its convolution with the lowpass filter (darkred). It detects a huge amount of false positives at segments with high conductance and high variance.}
\label{fig:PorBHeteroTRANSIT}
\end{figure}

\begin{figure}[!htb]
\centering
\includegraphics[width = 0.99\textwidth]{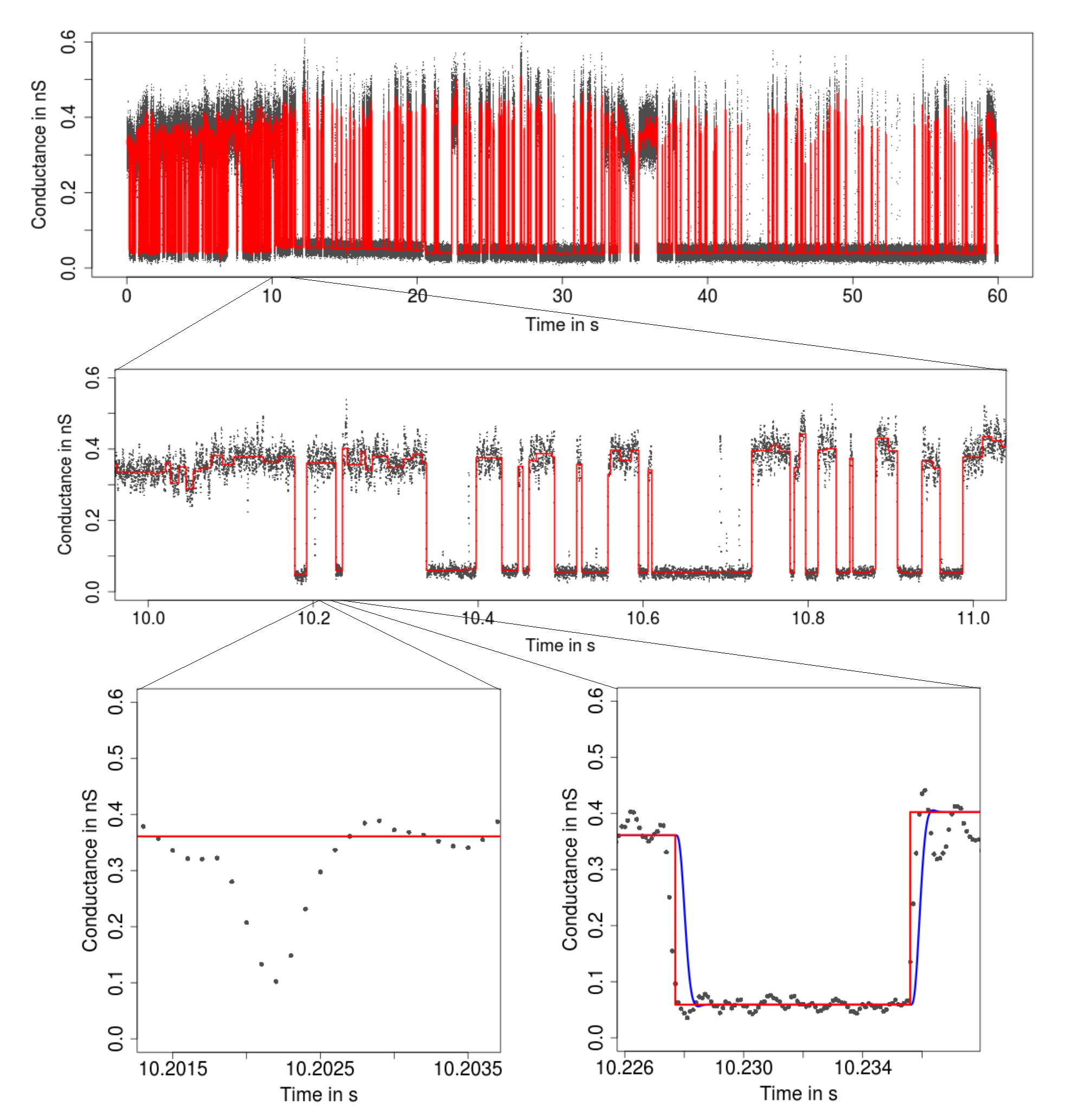}
\caption{Idealization (red) of the data in Figure \ref{fig:PorBHeteroData} by $\operatorname{HSMUCE}$ using a large significance of $\alpha = 0.99$ displayed on three different temporal scales. In the lower panels we also show the convolution of the idealization with the lowpass filter (blue). It still misses short events (for instance the one displayed in the lower left panel), but also includes several false positives.}
\label{fig:PorBHeteroHSMUCE99}
\end{figure}

\begin{figure}[!htb]
\centering
\includegraphics[width = 0.99\textwidth]{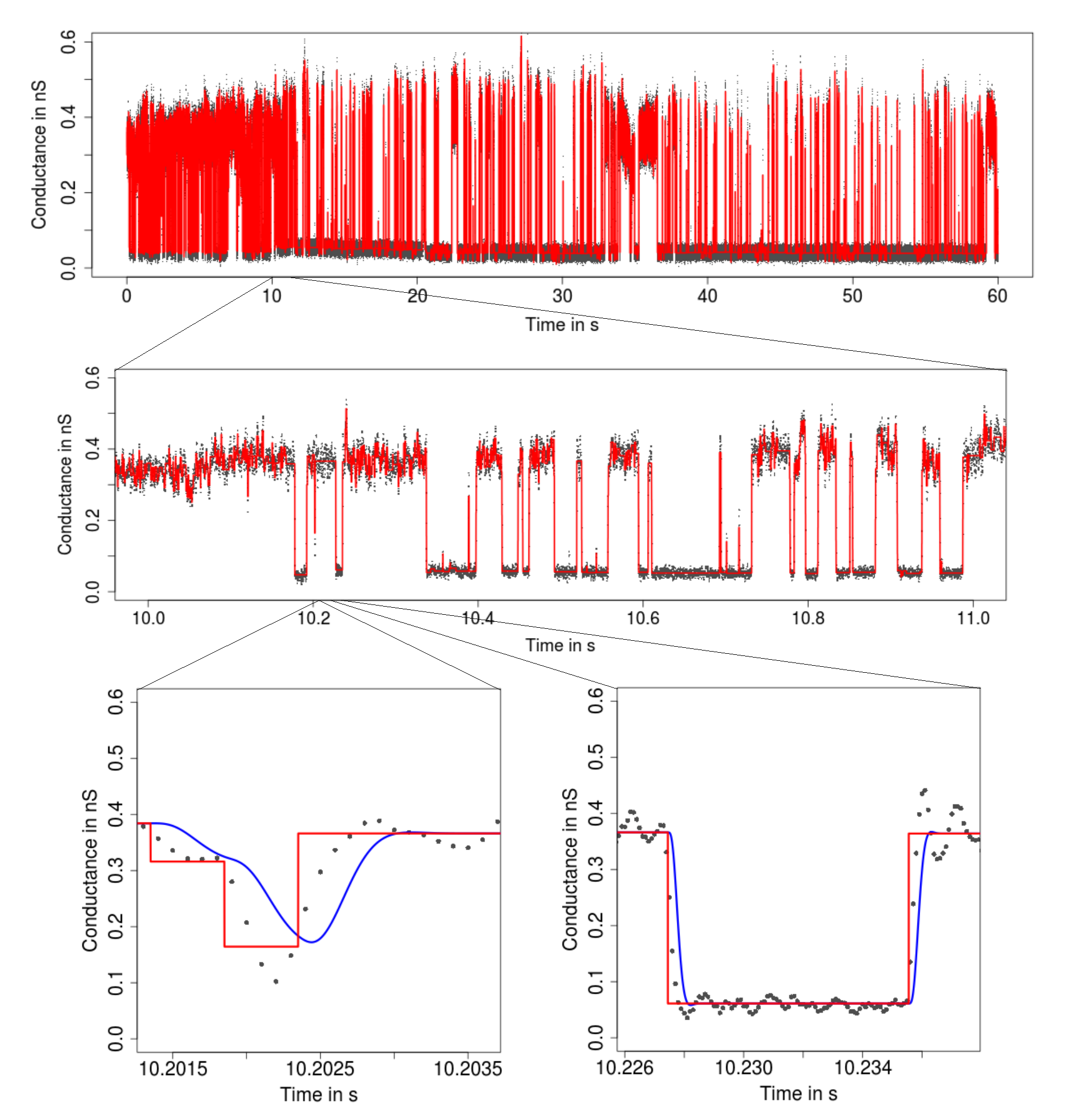}
\caption{Idealization (red) of the data in Figure \ref{fig:PorBHeteroData} by $\operatorname{CBS}$ displayed on three different temporal scales. In the lower panels we also show the convolution of the idealization with the lowpass filter (blue). It detects a huge amount of false positives at segments with high conductance and high variance.}
\label{fig:PorBHeteroCBS}
\end{figure}

It detects short events very well, for instance the events between \SI{10}{\second} and \SI{10.15}{\second} could be true events that are missed by other approaches. However, it also detects a huge amount of most likely false positives events at parts of the data with low conductivity, e.g., before \SI{10.4}{\second} (see for instance the lower right panel) and between \SI{10.6}{\second} and \SI{10.75}{\second}. It appears likely that these events are false positives, since there is not any visible indication of an event in the data at these locations. False positives can for instance be explained by missestimated parameters due to a too small number of states or also because of the previously highlighted artifacts in the data set.\\
Secondly, we discussed in Figure \ref{fig:PorBHeteroJULES} in the introduction an idealization by $\JULES$ \citep{Peinetal18}. More precisely, we used its implementation in R given by the function \textit{jules} in the CRAN package \textit{clampSeg} \citep{clampSeg} with its default parameter. This method is designed to take into account the filtering explicitly to idealize short events well, but assumes homogeneous noise. Hence, we found that it detects many small false positives on segments with a larger conductance and variance. Another model-free ion channel idealization approach is $\TRANSIT$ \citep{VanDongen.96}. Here, we used its implementation in R given by the function \textit{transit} in the CRAN package \textit{stepR} \citep{stepR} with its default parameter. Its idealization is shown in Figure \ref{fig:PorBHeteroTRANSIT}. Like $\JULES$ it detects many small false positives on segments with a larger conductance and variance. Such false positives are not a specific flaw of these methods, they will occur for any reasonable idealization method that ignores the heterogeneous noise.\\
Finally, we discussed in the introduction in Figure \ref{fig:PorBHeteroHSMUCE} an idealization by $\operatorname{HSMUCE}$, which serves as an example for a method that takes into account the heterogeneous noise, but ignores the filtering. We used the R function \textit{stepFit} of the CRAN package \textit{stepR} \citep{stepR} with the parametric family \textit{"hsmuce"} and a conservative significance level of $\alpha = 0.05$ to avoid overfitting. As discussed in the introduction, it works well on larger time scales, but is not able to detect shorter events. If we increase $\alpha$, this effect reduces, but at the precise of additional false positives, see Figure \ref{fig:PorBHeteroHSMUCE99}, where we used $\alpha = 0.99$. This effect is even more pronounced when we use $\operatorname{CBS}$ \citep{CBS04}, see Figure \ref{fig:PorBHeteroCBS}, a method that like $\operatorname{HSMUCE}$ takes into account heterogeneous noise, but ignores the filtering. However, in comparison to $\operatorname{HSMUCE}$ it puts less emphasize on avoiding false positives, but shows generally a higher detection power for very short events. In fact, we found that $\operatorname{CBS}$ is able to find short events, but at the price of a massive overfit. A last alternative could be subsampling to mitigate the filtering effects. However, this obviously also does not allow to detect short events and we did not display such results to avoid further lengthening of the paper.\\
In summary, none of the existing methods was able to idealize the data set in the middle panel of Figure \ref{fig:PorBHeteroData} reliably. In some cases, data cleaning or postfiltering might improve results, but this not only a huge amount of work that is required for every new data set again, it is also highly subjective. In comparison, $\HILDE$ provided in Figure \ref{fig:PorBHeteroJILTAD} a reasonable idealization without that anything like this was required.

\section{Analysis of flickering dynamics using idealizations of existing approaches}\label{sec:appendixanalysisothers}
\begin{figure}[!htb]
\centering
\begin{subfigure}{0.32\textwidth}
\includegraphics[width = \textwidth]{figures/Amplitude_PorBhetero_JILTAD.png}
\subcaption{$\HILDE$}
\label{subfig:PorBamplitudeHILDE}
\end{subfigure}
\begin{subfigure}{0.32\textwidth}
\includegraphics[width = \textwidth]{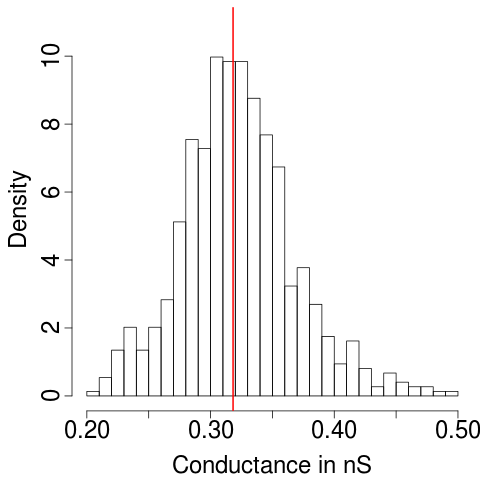}
\subcaption{$\JULES$}
\label{subfig:PorBamplitudeJULES}
\end{subfigure}
\begin{subfigure}{0.32\textwidth}
\includegraphics[width = \textwidth]{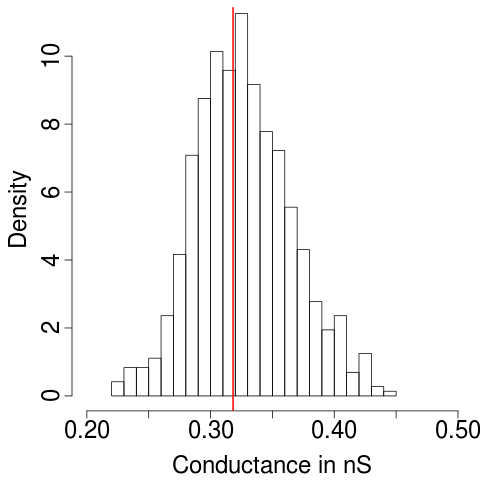}
\subcaption{$\HSMUCE$}
\label{subfig:PorBamplitudeHSMUCE}
\end{subfigure}
\caption{Histograms of the estimated amplitudes using various idealization approaches. The the red line indicates the estimated amplitudes of \SI{0.3194}{\nano\siemens}, \SI{0.3182}{\nano\siemens} and \SI{0.3216}{\nano\siemens} by the half sample mode using the idealizations from $\HILDE$, $\JULES$ and $\HSMUCE$, respectively.}
\label{fig:PorBOthersamplitude}
\end{figure}

In Section \ref{sec:analysisGating} we analyzed the gating dynamics of the PorB traces using the idealizations obtained by $\HILDE$. In this section we will examine whether we can obtain the same results when we use other idealization methods instead. Given the results from the previous section, we restrict ourself to $\JULES$, $\HSMUCE$ and the hidden Markov approach which assumes three states but with shared expectation and standard deviation for two states (in the following denoted by $\HMM$). Figure \ref{fig:PorBOthersamplitude} shows histograms of the amplitudes.

\begin{figure}[!htb]
\centering
\begin{subfigure}{0.24\textwidth}
\includegraphics[width = \textwidth]{figures/Dwell_PorBhetero_JILTAD.png}
\subcaption{$\HILDE$}
\label{subfig:PorBdwellHILDE}
\end{subfigure}
\begin{subfigure}{0.24\textwidth}
\includegraphics[width = \textwidth]{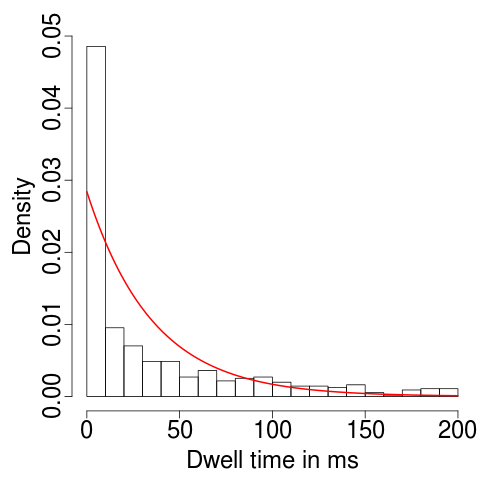}
\subcaption{$\JULES$}
\label{subfig:PorBdwellJULES}
\end{subfigure}
\begin{subfigure}{0.24\textwidth}
\includegraphics[width = \textwidth]{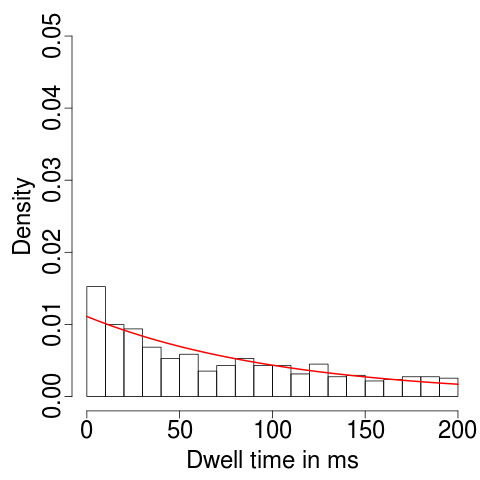}
\subcaption{$\HSMUCE$}
\label{subfig:PorBdwellHSMUCE}
\end{subfigure}
\begin{subfigure}{0.24\textwidth}
\includegraphics[width = \textwidth]{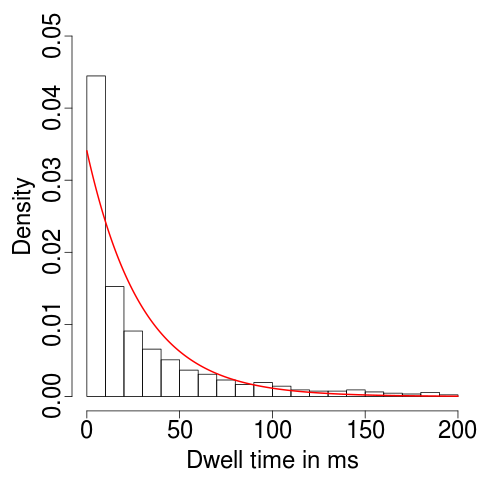}
\subcaption{$\HMM$}
\label{subfig:PorBdwellHMM}
\end{subfigure}
\caption{Histograms of the dwell times of opening events with amplitude between $\SI{0.2}{\nano\siemens}$ and $\SI{0.5}{\nano\siemens}$ together with exponential fits using a missed event correction (red line). We consider all dwell times between $\SI{0.1}{\milli\second}$ and $\SI{200}{\milli\second}$.}
\label{fig:PorBOthersdwell}
\end{figure}

\begin{figure}[!htb]
\centering
\begin{subfigure}{0.24\textwidth}
\includegraphics[width = \textwidth]{figures/DwellShort_PorBhetero_JILTAD.png}
\subcaption{$\HILDE$}
\label{subfig:PorBdwellShortHILDE}
\end{subfigure}
\begin{subfigure}{0.24\textwidth}
\includegraphics[width = \textwidth]{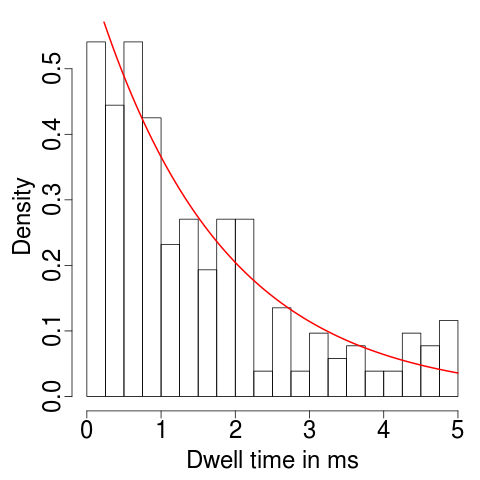}
\subcaption{$\JULES$}
\label{subfig:PorBdwellShortJULES}
\end{subfigure}
\begin{subfigure}{0.24\textwidth}
\includegraphics[width = \textwidth]{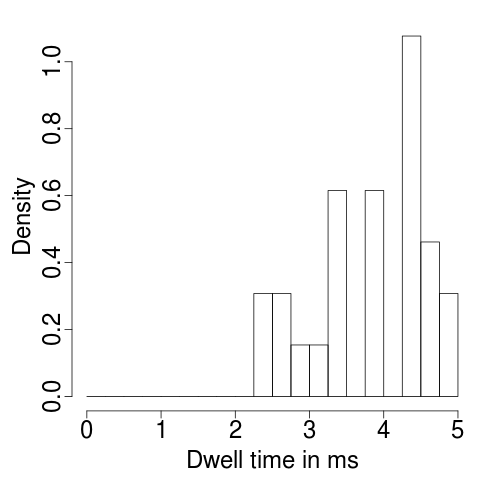}
\subcaption{$\HSMUCE$}
\label{subfig:PorBdwellShortHSMUCE}
\end{subfigure}
\begin{subfigure}{0.24\textwidth}
\includegraphics[width = \textwidth]{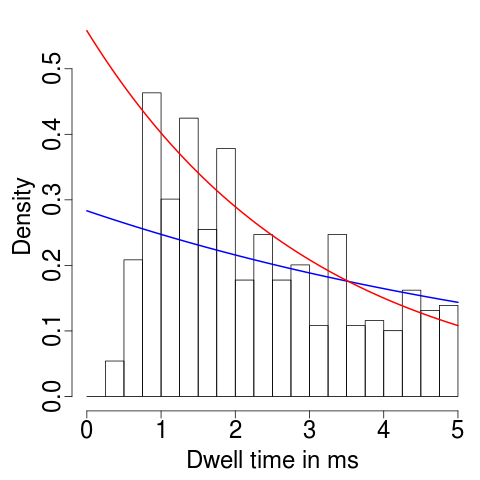}
\subcaption{$\HMM$}
\label{subfig:PorBdwellShortHMM}
\end{subfigure}
\caption{Histograms of the dwell times of opening events with amplitude between $\SI{0.2}{\nano\siemens}$ and $\SI{0.5}{\nano\siemens}$ together with exponential fits using a missed event correction (red line). We consider short dwell times between $\SI{0.1}{\milli\second}$ and $\SI{5}{\milli\second}$.}
\label{fig:PorBOthersdwellShort}
\end{figure}

\begin{figure}[!htb]
\centering
\begin{subfigure}{0.24\textwidth}
\includegraphics[width = \textwidth]{figures/DwellLong_PorBhetero_JILTAD.png}
\subcaption{$\HILDE$}
\label{subfig:PorBdwellLongHILDE}
\end{subfigure}
\begin{subfigure}{0.24\textwidth}
\includegraphics[width = \textwidth]{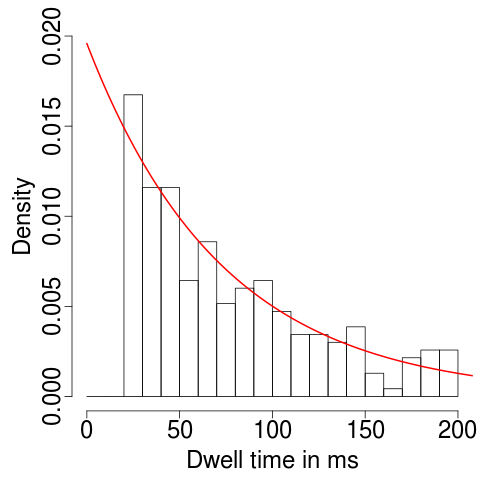}
\subcaption{$\JULES$}
\label{subfig:PorBdwellLongJULES}
\end{subfigure}
\begin{subfigure}{0.24\textwidth}
\includegraphics[width = \textwidth]{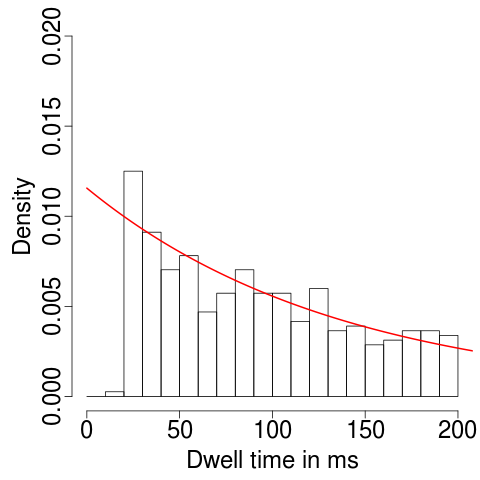}
\subcaption{$\HSMUCE$}
\label{subfig:PorBdwellLongHSMUCE}
\end{subfigure}
\begin{subfigure}{0.24\textwidth}
\includegraphics[width = \textwidth]{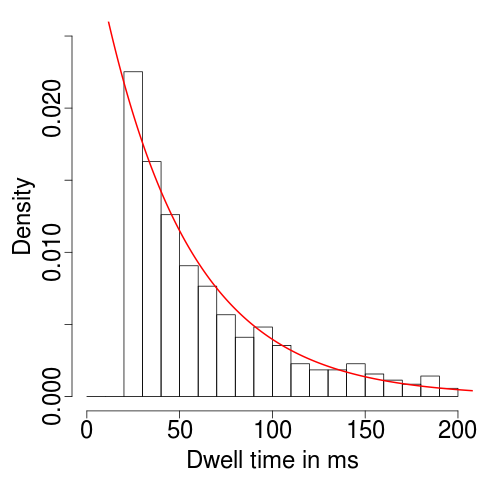}
\subcaption{$\HMM$}
\label{subfig:PorBdwellLongHMM}
\end{subfigure}
\caption{Histograms of the dwell times of opening events with amplitude between $\SI{0.2}{\nano\siemens}$ and $\SI{0.5}{\nano\siemens}$ together with exponential fits using a missed event correction (red line). We consider long dwell times between $\SI{20}{\milli\second}$ and $\SI{200}{\milli\second}$.}
\label{fig:PorBOthersdwellLong}
\end{figure}

We note that all model-free approaches estimate roughly the same amplitude. However, when we used the Baum-Welch algorithm to fit the hidden Markov model we obtained an amplitude of \SI{0.8124}{\nano\siemens} which is far off and most likely caused by artifacts. Next we consider in Figures \ref{fig:PorBOthersdwell}-\ref{fig:PorBOthersdwellLong} the estimated dwell times in the open state.

\begin{figure}[!htb]
\centering
\begin{subfigure}{0.24\textwidth}
\includegraphics[width = \textwidth]{figures/Distance_PorBhetero_JILTAD.png}
\subcaption{$\HILDE$}
\label{subfig:PorBdistanceHILDE}
\end{subfigure}
\begin{subfigure}{0.24\textwidth}
\includegraphics[width = \textwidth]{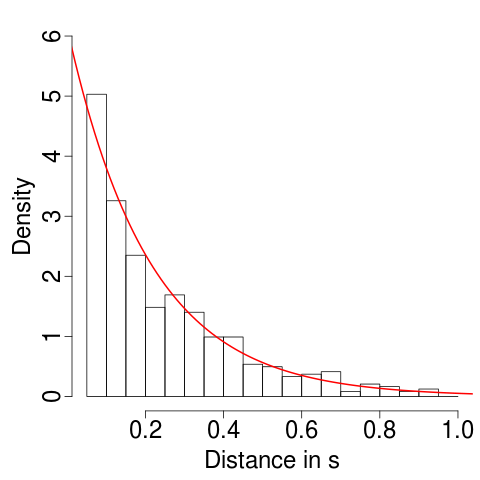}
\subcaption{$\JULES$}
\label{subfig:PorBdistanceJULES}
\end{subfigure}
\begin{subfigure}{0.24\textwidth}
\includegraphics[width = \textwidth]{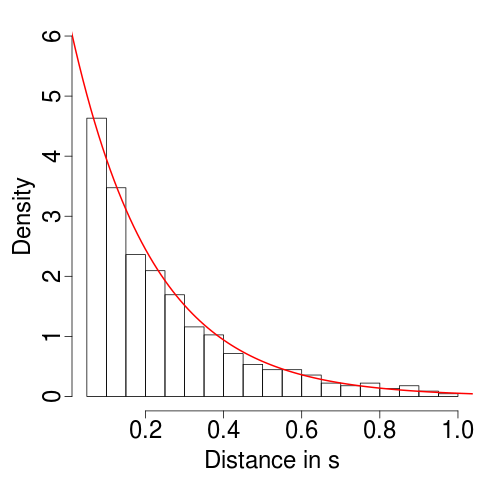}
\subcaption{$\HSMUCE$}
\label{subfig:PorBdistanceHSMUCE}
\end{subfigure}
\begin{subfigure}{0.24\textwidth}
\includegraphics[width = \textwidth]{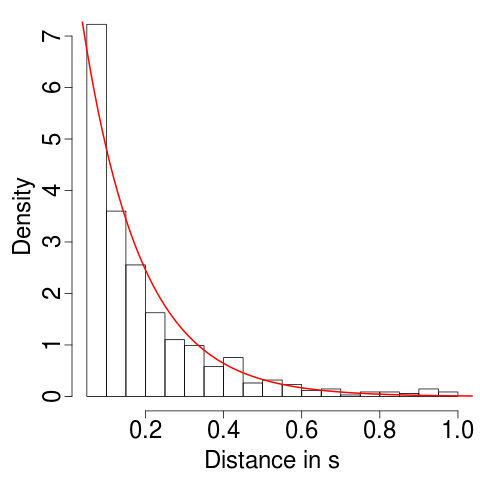}
\subcaption{$\HMM$}
\label{subfig:PorBdistanceHMM}
\end{subfigure}
\caption{Histograms of the dwell times in the closed state, together with exponential fits (red) that are corrected for missed events.}
\label{fig:PorBothersDistance}
\end{figure}

We found in Figures \ref{fig:PorBOthersdwell}-\ref{fig:PorBOthersdwellLong} that all comparisons are qualitatively the same as for the hidden Markov model simulations in Section \ref{sec:hmm}. Once again $\HSMUCE$ is not able to detect short events and $\HMM$ requires a stricter missed event analysis to analyze them well. Using $\HILDE$, $\JULES$ and $\HMM$ we obtained an average duration of \SI{2.31}{\milli\second}, \SI{1.72}{\milli\second} and \SI{3.05}{\milli\second} for the short events and using $\HILDE$, $\JULES$, $\HSMUCE$ and $\HMM$ we obtained an average duration of \SI{51.62}{\milli\second}, \SI{73.37}{\milli\second}, \SI{136.85}{\milli\second} and \SI{46.96}{\milli\second} for the long events. Hence, the estimates obtained by $\HILDE$ are roughly confirmed. Finally, we are now analyzing the dwell times in the closed state (Figure \ref{fig:PorBothersDistance}) and the proportions of short and long events.

We found in Figure \ref{fig:PorBothersDistance} that all four methods indentified an exponential distrubtion and also the estimate rates are rather similar: Using $\HILDE$, $\JULES$, $\HSMUCE$ and $\HMM$ we obtained a frequency of \SI{5.75}{\hertz}, \SI{4.80}{\hertz}, \SI{5.66}{\hertz} and \SI{6,74}{\hertz}, respectively. And using $\HILDE$, $\JULES$, and $\HMM$ we estimated the proportion of short events to be $39.08\%$, $40.79\%$ and $44.36\%$, respectively. In summary, all results obtained by using $\HILDE$ could be confirmed by at least one other approach. However, one should also note that none of the other methods was able to reproduce all results obtained by $\HILDE$. This is confirmed by simulations in Section \ref{sec:hmm}, where we simulated data from a hidden Markov similar to the one we estimated from the PorB data.

\section{Homogeneous noise}\label{sec:appendixhomogeneousnoise}
This section details how $\HILDE$ can be adapted to the assumption of homogeneous noise. A constant variance can be preestimated as in (6) in \citep{Pein.Sieling.Munk.16}. The first multiresolution step to detect events on larger scales can be performed by $\JSMURF$ as defined in \citep{Hotz.etal.13}. And for the local tests to detect events on smaller scales in the second step we suggest the (regularized) likelihood ratio test statistic
\begin{equation*}
\begin{split}
T_i^j:=&\left(Y_{i+1,j+m-1}-(\valmu_0)_{i+1,j+m-1}\right)^t \Sigma_{i+1,j+m-1}^{-1}\left(Y_{i+1,j+m-1}-(\valmu_0)_{i+1,j+m-1}\right)\\
- &\left(Y_{i+1,j+m-1}-(\valmu_1(\hat{\valmu}))_{i+1,j+m-1}\right)^t \Sigma_{i+1,j+m-1}^{-1}\left(Y_{i+1,j+m-1}-(\valmu_1(\hat{\valmu}))_{i+1,j+m-1}\right),
\end{split}
\end{equation*}
with 
\begin{equation*}
\hat{\valmu}:=\operatorname{argmax}_{\valmu\in \R}{\left(Y_{i+1,j+m-1}-(\valmu_1(\valmu))_{i+1,j+m-1}\right)^t \Sigma_{i+1,j+m-1}^{-1}\left(Y_{i+1,j+m-1}-(\valmu_1(\valmu))_{i+1,j+m-1}\right)}.
\end{equation*}
Here, $(\valmu_0)_{i+1,j+m-1}$ and $(\valmu_1(\valmu))_{i+1,j+m-1}$ are the vectors
\begin{equation*}
\begin{split}
&(\valmu_0)_{i+1,j+m-1}=\big((\kernel_m\ast f_0)((i + 1)/\sr),\ldots,(\kernel_m\ast f_0)((j+m-1)/\sr)\big)^t,\\
&(\valmu_1(\valmu))_{i+1,j+m-1}=\big((\kernel_m\ast f_1(\valmu))((i + 1)/\sr),\ldots,(\kernel_m\ast f_1(\valmu))((j+m-1)/\sr)\big)^t
\end{split}
\end{equation*}
and $\Sigma_{i+1,j+m-1}$ the covariance matrix of the observations $Y_{i+1},\ldots,Y_{j+m-1}$ given by \eqref{eq:covpeak} and regularized by Tikhonov regularization with parameter $\gamma^2=\sigma_0^2$. And, recall, $Y_{i,j}=(Y_i,\ldots,Y_j)^t$ and $f_0$ and $f_1$ are given in \eqref{eq:hypothesis} and \eqref{eq:alternative}, respectively. 

\end{document}